\documentclass[pdftex,twocolumn,epjc3_preprint,runningheads]{svjour3}
\pdfoutput=1

\usepackage[T1]{fontenc}
\usepackage{lmodern}
\usepackage{calc}
\usepackage{graphicx}
\usepackage{booktabs}
\usepackage{textcomp}
\usepackage{xspace}
\usepackage{relsize}
\usepackage{amssymb}
\usepackage{amsmath}
\usepackage{listings}
\usepackage{microtype}
\usepackage{multirow}
\usepackage{tabularx}
\usepackage{array}
\usepackage{placeins}
\usepackage{cuted}
\usepackage{soul} 
\usepackage{fixltx2e}
\usepackage[numbers,sort&compress]{natbib}
\usepackage[labelfont=bf,font=small]{caption}
\usepackage[skip=-2pt]{subcaption}
\usepackage[colorlinks,citecolor=blue,urlcolor=blue,linkcolor=blue,breaklinks=breakall]{hyperref}
\usepackage{breakurl}
\usepackage[dvipsnames]{xcolor}
\usepackage[clockwise,figuresright]{rotating}
\usepackage{siunitx}
\usepackage{tikz}

\usepackage{etoolbox}
\AfterEndEnvironment{strip}{\leavevmode}

\allowdisplaybreaks

\newcolumntype{L}{>{\raggedright\let\newline\\\arraybackslash\hspace{0pt}}X}
\newcolumntype{R}{>{\raggedleft\let\newline\\\arraybackslash\hspace{0pt}}X}
\newcolumntype{C}{>{\centering\let\newline\\\arraybackslash\hspace{0pt}}X}

\newcommand{\imperial}{Department of Physics, Imperial College London, Blackett Laboratory, Prince Consort Road, London SW7 2AZ, UK}
\newcommand{\nordita}{NORDITA, Roslagstullsbacken 23, SE-10691 Stockholm, Sweden}
\newcommand{\oslo}{Department of Physics, University of Oslo, N-0316 Oslo, Norway}
\newcommand{\adelaide}{Department of Physics, University of Adelaide, Adelaide, SA 5005, Australia}

\newcommand{\coepp}{Australian Research Council Centre of Excellence for Particle Physics at the Tera-scale}
\newcommand{\okc}{Oskar Klein Centre for Cosmoparticle Physics, AlbaNova University Centre, SE-10691 Stockholm, Sweden}
\newcommand{\su}{Department of Physics, Stockholm University, SE-10691 Stockholm, Sweden}

\newcommand{\annecy}{LAPTh, Universit\'e de Savoie, CNRS, 9 chemin de Bellevue B.P.110, F-74941 Annecy-le-Vieux, France}
\newcommand{\harvard}{Department of Physics, Harvard University, Cambridge, MA 02138, USA}
\newcommand{\grappa}{GRAPPA, Institute of Physics, University of Amsterdam, Science Park 904, 1098 XH Amsterdam, Netherlands}

\newcommand{\cernth}{Theoretical Physics Department, CERN, CH-1211 Geneva 23, Switzerland}
\newcommand{\lyon}{Univ Lyon, Univ Lyon 1, ENS de Lyon, CNRS, Centre de Recherche Astrophysique de Lyon UMR5574, F-69230 Saint-Genis-Laval, France}
\newcommand{\iuf}{Institut Universitaire de France, 103 boulevard Saint-Michel, 75005 Paris, France}
\newcommand{\zurich}{Physik-Institut, Universit\"at Z\"urich, Winterthurerstrasse 190, 8057 Z\"urich, Switzerland}
\newcommand{\krakow}{H.~Niewodnicza\'nski Institute of Nuclear Physics, Polish Academy of Sciences, 31-342  Krak\'ow, Poland}
\newcommand{\bonn}{Physikalisches Institut der Rheinischen Friedrich-Wilhelms-Universit\"at Bonn, 53115 Bonn, Germany}

\newcommand{\gambitacknos    }{We warmly thank the Casa Matem\'aticas Oaxaca, affiliated with the Banff International Research Station, for hospitality whilst part of this work was completed, and the staff at Cyfronet, for their always helpful supercomputing support.  \GB has been supported by STFC (UK; ST/K00414X/1, ST/P000762/1), the Royal Society (UK; UF110191), Glasgow University (UK; Leadership Fellowship), the Research Council of Norway (FRIPRO 230546/F20), NOTUR (Norway; NN9284K), the Knut and Alice Wallenberg Foundation (Sweden; Wallenberg Academy Fellowship), the Swedish Research Council (621-2014-5772), the Australian Research Council (CE110001004, FT130100018, FT140100244, FT160100274), The University of Sydney (Australia; IRCA-G162448), PLGrid Infrastructure (Poland), Polish National Science Center (Sonata UMO-2015/17/D/ST2/03532), the Swiss National Science Foundation (PP00P2-144674), the European Commission Horizon 2020 Marie Sk\l{}odowska-Curie actions (H2020-MSCA-RISE-2015-691164), the ERA-CAN+ Twinning Program (EU \& Canada), the Netherlands Organisation for Scientific Research (NWO-Vidi 680-47-532), the National Science Foundation (USA; DGE-1339067), the FRQNT (Qu\'ebec) and NSERC/The Canadian Tri-Agencies Research Councils (BPDF-424460-2012).}

\makeatletter

\newcommand{\preprintnumber}[1]{\gdef\@preprintnumber{\begin{flushright}{#1}\end{flushright}}}

\g@addto@macro\bfseries{\boldmath}
\makeatother

\newcommand{\subparagraph}{} 
\usepackage{titlesec}
\titleformat*{\paragraph}{\bfseries}

\journalname{Eur. Phys. J. C}
\smartqed
\let\underscore\_
\renewcommand{\_}{\discretionary{\underscore}{}{\underscore}}

\makeatletter
\let\orgdescriptionlabel\descriptionlabel
\renewcommand*{\descriptionlabel}[1]{%
  \let\orglabel\label
  \let\label\@gobble
  \phantomsection
  \protected@edef\@currentlabel{#1}%
  \let\label\orglabel
  \orgdescriptionlabel{#1}%
}
\makeatother

\newcommand\postnewlinemarker{\hbox{\ensuremath{\hookrightarrow}}}
\newcommand\cpp[1]{{\lstinline!#1!}}  

\newcommand\yaml[1]{{\lstset{style=yaml}\lstinline!#1!\lstset{style=cpp}}}
\newcommand\yamlvalue[1]{{\YAMLvaluestyle\ttfamily#1}}
\newcommand\term[1]{{\lstset{style=terminal}\lstinline!#1!\lstset{style=cpp}}}
\newcommand\fortran[1]{{\lstset{style=fortran}\lstinline!#1!\lstset{style=cpp}}}
\newcommand\py[1]{{\lstset{style=python}\lstinline!#1!\lstset{style=cpp}}}
\newcommand\customtilde{{\raisebox{0.2ex}{\scalebox{0.6}{\boldmath$\sim$}}}}
\newcommand\mathematica[1]{{\lstset{style=Mathematica}\lstinline!#1!\lstset{style=cpp}}}

\lstnewenvironment{lstlistingyaml}{\lstset{style=yaml}}{\lstset{style=cpp}}
\lstnewenvironment{lstlistingterm}{\lstset{style=terminal}}{\lstset{style=cpp}}
\lstnewenvironment{lstlistingfortran}{\lstset{style=fortran}}{\lstset{style=cpp}}
\lstnewenvironment{lstcpp}{\lstset{style=cpp}}{\lstset{style=cpp}}
\lstnewenvironment{lstcppalt}{\lstset{style=cppalt}}{\lstset{style=cpp}}
\lstnewenvironment{lstcppnum}{\lstset{style=cppnum}}{\lstset{style=cpp}}
\lstnewenvironment{lstyaml}{\lstset{style=yaml}}{\lstset{style=cpp}}
\lstnewenvironment{lstterm}{\lstset{style=terminal}}{\lstset{style=cpp}}
\lstnewenvironment{lsttermalt}{\lstset{style=terminalalt}}{\lstset{style=cpp}}
\lstnewenvironment{lsttext}{\lstset{style=text}}{\lstset{style=cpp}}
\lstnewenvironment{lstfortran}{\lstset{style=fortran}}{\lstset{style=cpp}}
\lstnewenvironment{lstpy}{\lstset{style=python}}{\lstset{style=cpp}}
\lstnewenvironment{lstmathematica}{\lstset{style=mathematica}}{\lstset{style=cpp}}

\newcommand{\tmpname}{}
\newcommand{\tmplistingname}{}
\makeatletter
\newif\ifATOlabelname
\lst@Key{labelname}{Listing}{\def\ATOlabelname{#1}\global\ATOlabelnametrue}
\makeatother
\lstnewenvironment{lstcpplabel}[1][]{
  \lstset{style=cpp,#1} 
  \ifATOlabelname
    \renewcommand{\tmpname}{\lstlistingname}
    \renewcommand{\tmplistingname}{\lstlistlistingname}
    \renewcommand{\lstlistingname}{\ATOlabelname}
    \renewcommand{\lstlistlistingname}{List of \lstlistingname s}
  \fi
}{
  \renewcommand{\lstlistingname}{\tmpname}
  \renewcommand{\lstlistlistingname}{\tmplistingname}
  \lstset{style=cpp}
}
\definecolor{solarized@base03}{HTML}{002B36}
\definecolor{solarized@base02}{HTML}{073642}
\definecolor{solarized@base01}{HTML}{586e75}
\definecolor{solarized@base00}{HTML}{657b83}
\definecolor{solarized@base0}{HTML}{839496}
\definecolor{solarized@base1}{HTML}{93a1a1}
\definecolor{solarized@base2}{HTML}{EEE8D5}
\definecolor{solarized@base3}{HTML}{FDF6E3}
\definecolor{solarized@yellow}{HTML}{B58900}
\definecolor{solarized@orange}{HTML}{CB4B16}
\definecolor{solarized@red}{HTML}{DC322F}
\definecolor{solarized@magenta}{HTML}{D33682}
\definecolor{solarized@violet}{HTML}{6C71C4}
\definecolor{solarized@blue}{HTML}{268BD2}
\definecolor{solarized@cyan}{HTML}{2AA198}
\definecolor{solarized@green}{HTML}{859900}
\definecolor{darkred}{HTML}{550003}
\definecolor{darkgreen}{HTML}{00AA00}

\newcommand\YAMLstringstyle{\footnotesize\color{solarized@green}\mdseries}
\newcommand\YAMLkeystyle{\footnotesize\color{solarized@blue}\ttfamily}
\newcommand\YAMLvaluestyle{\footnotesize\color{blue}\mdseries}
\newcommand\ProcessThreeDashes{\llap{\color{cyan}\mdseries-{-}-}}

\newcommand\CPPcommentstyle{\color{solarized@violet}\footnotesize\ttfamily}
\newcommand\CPPdirectivestyle{\color{solarized@magenta}\footnotesize\ttfamily}
\newcommand\termplainstyle{\footnotesize\ttfamily}

\newcommand\processLongMacroDelimiter
{%
\CPPdirectivestyle%
\#define%
}

\lstdefinestyle{cpp}
{
  language=C++,
  basicstyle=\footnotesize\ttfamily,
  basewidth={0.53em,0.44em}, 
  numbers=none,
  tabsize=2,
  breaklines=true,
  escapeinside={@}{@},
  showstringspaces=false,
  numberstyle=\tiny\color{solarized@base01},
  keywordstyle=\color{solarized@orange},
  stringstyle=\color{solarized@red}\ttfamily,
  identifierstyle=\color{solarized@blue},
  commentstyle=\CPPcommentstyle,
  directivestyle=\CPPdirectivestyle,
  emphstyle=\color{solarized@green},
  frame=single,
  rulecolor=\color{solarized@base2},
  rulesepcolor=\color{solarized@base2},
  literate={~} {\customtilde}1,
  moredelim=*[directive]\ \ \#,
  moredelim=*[directive]\ \ \ \ \#
}

\lstdefinestyle{cppalt}
{
  language=C++,
  basicstyle=\footnotesize\ttfamily,
  basewidth={0.53em,0.44em}, 
  numbers=none,
  tabsize=2,
  breaklines=true,
  escapeinside={*@}{@*},
  showstringspaces=false,
  numberstyle=\tiny\color{solarized@base01},
  keywordstyle=\color{solarized@orange},
  stringstyle=\color{solarized@red}\ttfamily,
  identifierstyle=\color{solarized@blue},
  commentstyle=\CPPcommentstyle,
  directivestyle=\CPPdirectivestyle,
  emphstyle=\color{solarized@green},
  frame=single,
  rulecolor=\color{solarized@base2},
  rulesepcolor=\color{solarized@base2},
  literate={~}{\customtilde}1,
  moredelim=**[is][\processLongMacroDelimiter]{BeginLongMacro}{EndLongMacro} 
}

\lstdefinestyle{cppnum}
{
  language=C++,
  basicstyle=\footnotesize\ttfamily,
  basewidth={0.53em,0.44em}, 
  numbers=none,
  tabsize=2,
  breaklines=true,
  escapeinside={@}{@},
  numberstyle=\tiny\color{solarized@base01},
  showstringspaces=false,
  numberstyle=\tiny\color{solarized@base01},
  keywordstyle=\color{solarized@orange},
  stringstyle=\color{solarized@red}\ttfamily,
  identifierstyle=\color{solarized@blue},
  commentstyle=\CPPcommentstyle,
  directivestyle=\CPPdirectivestyle,
  emphstyle=\color{solarized@green},
  frame=single,
  rulecolor=\color{solarized@base2},
  rulesepcolor=\color{solarized@base2},
  literate={~} {\customtilde}1,
  moredelim=*[directive]\ \ \#,
  moredelim=*[directive]\ \ \ \ \#
}

\lstdefinestyle{python}
{
  language=Python,
  basicstyle=\footnotesize\ttfamily,
  basewidth={0.53em,0.44em},
  numbers=none,
  tabsize=2,
  breaklines=true,
  escapeinside={@}{@},
  showstringspaces=false,
  numberstyle=\tiny\color{solarized@base01},
  keywordstyle=\color{blue},
  stringstyle=\color{orange}\ttfamily,
  identifierstyle=\color{darkred},
  commentstyle=\color{purple},
  emphstyle=\color{green},
  frame=single,
  rulecolor=\color{solarized@base2},
  rulesepcolor=\color{solarized@base2},
  literate = {~}{\customtilde}1
             {\ as\ }{{\color{blue}\ as\ \color{black}}}3
}

\lstdefinestyle{fortran}
{
  language=Fortran,
  basicstyle=\footnotesize\ttfamily,
  basewidth={0.53em,0.44em},
  numbers=none,
  tabsize=2,
  breaklines=true,
  escapeinside={@}{@},
  showstringspaces=false,
  numberstyle=\tiny\color{solarized@base01},
  keywordstyle=\color{blue},
  stringstyle=\color{orange}\ttfamily,
  identifierstyle=\color{Periwinkle},
  commentstyle=\color{purple},
  emphstyle=\color{green},
  morekeywords={and, or, true, false},
  frame=single,
  rulecolor=\color{solarized@base2},
  rulesepcolor=\color{solarized@base2},
  literate={~}{\customtilde}1
}

\lstdefinestyle{terminal}
{
  language=bash,
  basicstyle=\termplainstyle,
  numbers=none,
  tabsize=2,
  breaklines=true,
  escapeinside={@}{@},
  frame=single,
  showstringspaces=false,
  numberstyle=\tiny\color{solarized@base01},
  keywordstyle=\color{solarized@orange},
  stringstyle=\color{solarized@red}\ttfamily,
  identifierstyle=\color{black},
  commentstyle=\color{solarized@violet},
  emphstyle=\color{solarized@green},
  frame=single,
  rulecolor=\color{solarized@base2},
  rulesepcolor=\color{solarized@base2},
  morekeywords={gambit, cmake, make, mkdir},
  deletekeywords={test},
  literate = {\ gambit}{{\ }{\color{black}}gambit}7
             {/gambit}{{/}{\color{black}}gambit}6
             {gambit/}{{\color{black}}gambit{/}}6
             {/include}{{/}{\color{black}}include}8
             {cmake/}{{\color{black}}cmake/}6
             {.cmake}{{.}{\color{black}}cmake}6
             {~}{\customtilde}1
}

\lstdefinestyle{terminalalt}
{
  language=bash,
  basicstyle=\footnotesize\ttfamily,
  numbers=none,
  tabsize=2,
  breaklines=true,
  escapeinside={*@}{@*},
  frame=single,
  showstringspaces=false,
  numberstyle=\tiny\color{solarized@base01},
  keywordstyle=\color{solarized@orange},
  stringstyle=\color{solarized@red}\ttfamily,
  identifierstyle=\color{black},
  commentstyle=\color{solarized@violet},
  emphstyle=\color{solarized@green},
  frame=single,
  rulecolor=\color{solarized@base2},
  rulesepcolor=\color{solarized@base2},
  morekeywords={gambit, cmake, make, mkdir},
  deletekeywords={test},
  literate = {\ gambit}{{\ }{\color{black}}gambit}7
             {/gambit}{{/}{\color{black}}gambit}6
             {gambit/}{{\color{black}}gambit{/}}6
             {/include}{{/}{\color{black}}include}8
             {cmake/}{{\color{black}}cmake/}6
             {.cmake}{{.}{\color{black}}cmake}6
             {~}{\customtilde}1
}

\lstdefinestyle{text}
{
  language={},
  basicstyle=\footnotesize\ttfamily,
  identifierstyle=\color{black},
  numbers=none,
  tabsize=2,
  breaklines=true,
  escapeinside={*@}{@*},
  showstringspaces=false,
  frame=single,
  rulecolor=\color{solarized@base2},
  rulesepcolor=\color{solarized@base2},
  literate={~}{\customtilde}1
}

\lstdefinestyle{yaml}
{
  language=bash,
  escapeinside={@}{@},
  keywords={true,false,null},
  otherkeywords={},
  keywordstyle=\color{solarized@base0}\bfseries,
  basicstyle=\footnotesize\color{black}\ttfamily,
  identifierstyle=\YAMLkeystyle,
  sensitive=false,
  commentstyle=\color{solarized@orange}\ttfamily,
  morecomment=[l]{\#},
  morecomment=[s]{/*}{*/},
  stringstyle=\YAMLstringstyle\ttfamily,
  moredelim=**[s][\YAMLkeystyle]{,}{:},   
  moredelim=**[l][\YAMLvaluestyle]{:},    
  morestring=[b]',
  morestring=[b]",
  literate =    {---}{{\ProcessThreeDashes}}3
                {>}{{\textcolor{solarized@red}\textgreater}}1
                {|}{{\textcolor{solarized@red}\textbar}}1
                {\ -\ }{{\mdseries\color{black}\ -\ \negmedspace}}3
                {\}}{{{\color{black} \}}}}1
                {\{}{{{\color{black} \{}}}1
                {[}{{{\color{black} [}}}1
                {]}{{{\color{black} ]}}}1
                {~}{\customtilde}1,
  breakindent=0pt,
  breakatwhitespace,
  columns=fullflexible
}

\lstdefinestyle{mathematica}
{
  language={Mathematica},
  basicstyle=\footnotesize\ttfamily,
  basewidth={0.53em,0.44em},
  numbers=none,
  tabsize=2,
  breaklines=true,
  escapeinside={@}{@},
  numberstyle=\tiny\color{black},
  showstringspaces=false,
  numberstyle=\tiny\color{solarized@base01},
  keywordstyle=\color{solarized@orange},
  stringstyle=\color{solarized@red}\ttfamily,
  identifierstyle=\color{solarized@orange}\ttfamily,
  commentstyle=\color{solarized@gray}\ttfamily,
  directivestyle=\color{solarized@orange}\ttfamily,
  emphstyle=\color{solarized@green},
  frame=single,
  rulecolor=\color{solarized@base2},
  rulesepcolor=\color{solarized@base2},
  literate={~} {\customtilde}1,
  moredelim=*[directive]\ \ \#,
  moredelim=*[directive]\ \ \ \ \#,
  mathescape=true
}

\newcommand{\cross}[1]{\ref{#1}}
\newcommand{\doublecross}[2]{\hyperref[#2]{\textbf{#1}}}
\newcommand{\doublecrosssf}[2]{\hyperref[#2]{\textbf{\textsf{#1}}}}
\newcommand{\gitem}[1]{\item[\textbf{#1}\label{#1}]}

\newcommand{\startglossary}{\section{Glossary}\label{glossary}Here we explain some terms that have specific technical definitions in \GB.\begin{description}}
\newcommand{\finishglossary}{\end{description}}

\newcommand{\bcode}{\begin{lstlisting}}
\newcommand{\ecode}{\end{lstlisting}}
\newcommand{\metavarf}[1]{\textit{\color{darkgreen}\footnotesize\textrm{#1}}}

\newcommand{\metavar}{\metavarf}

\newcommand{\DR}{$\overline{DR}$\xspace}
\newcommand{\DRbar}{\DR}
\newcommand{\MSbar}{$\MSBar$\xspace}
\newcommand{\MSBar}{\overline{MS}}

\newcommand{\gambit}{\textsf{GAMBIT}\xspace}

\newcommand{\flavbit}{\textsf{FlavBit}\xspace}
\newcommand{\specbit}{\textsf{SpecBit}\xspace}

\newcommand{\precisionbit}{\textsf{PrecisionBit}\xspace}

\newcommand{\GB}{\gambit}

\newcommand{\feynhiggs}{\textsf{FeynHiggs}\xspace}
\newcommand{\FH}{\feynhiggs}

\newcommand\flexiblesusy{\FlexibleSUSY}
\newcommand\FlexibleSUSY{\textsf{FlexibleSUSY}\xspace}

\newcommand\SOFTSUSY{\textsf{SOFTSUSY}\xspace}

\newcommand\superiso{\textsf{SuperIso}\xspace}

\newcommand\diver{\textsf{Diver}\xspace}

\newcommand\xx{\raisebox{0.2ex}{\smaller ++}\xspace}
\newcommand\Cpp{\textsf{C\xx}\xspace}

\newcommand\YAML{\textsf{YAML}\xspace}

\newcommand\beq{\begin{equation}}
\newcommand\eeq{\end{equation}}

\renewcommand{\url}[1]{\href{#1}{#1}}

\def\MagUp {\mbox{\em Mag\kern -0.05em Up}\xspace}

 \def\PDelta      {\ensuremath{\Delta}\xspace}
 \def\PXi      {\ensuremath{\Xi}\xspace}
 \def\PLambda      {\ensuremath{\Lambda}\xspace}
 \def\PSigma      {\ensuremath{\Sigma}\xspace}
 \def\POmega      {\ensuremath{\Omega}\xspace}
 \def\PUpsilon      {\ensuremath{\Upsilon}\xspace}

 \def\PB      {\ensuremath{\mathrm{B}}\xspace}
 
 \def\PD      {\ensuremath{\mathrm{D}}\xspace}

 \def\PK      {\ensuremath{\mathrm{K}}\xspace}

 \def\Pi      {\ensuremath{\mathrm{i}}\xspace}

{

 \mathchardef\PDelta="7101
 \mathchardef\PXi="7104
 \mathchardef\PLambda="7103
 \mathchardef\PSigma="7106
 \mathchardef\POmega="710A
 \mathchardef\PUpsilon="7107
 
 \def\PB      {\ensuremath{B}\xspace}
 
 \def\PD      {\ensuremath{D}\xspace}

 \def\PK      {\ensuremath{K}\xspace}

 \def\Pi      {\ensuremath{i}\xspace}

}

\makeatletter
\ifcase \@ptsize \relax
  \newcommand{\miniscule}{\@setfontsize\miniscule{4}{5}}
\or
  \newcommand{\miniscule}{\@setfontsize\miniscule{5}{6}}
\or
  \newcommand{\miniscule}{\@setfontsize\miniscule{5}{6}}
\fi
\makeatother

\DeclareRobustCommand{\optbar}[1]{\shortstack{{\miniscule (\rule[.5ex]{1.25em}{.18mm})}
  \\ [-.7ex] $#1$}}

  \def\Kbar    {{\kern 0.2em\overline{\kern -0.2em \PK}{}}\xspace}

\def\KorKbar    {\kern 0.18em\optbar{\kern -0.18em K}{}\xspace}

  \def\Dbar    {{\kern 0.2em\overline{\kern -0.2em \PD}{}}\xspace}

\def\DorDbar    {\kern 0.18em\optbar{\kern -0.18em D}{}\xspace}

\def\Bbar    {{\ensuremath{\kern 0.18em\overline{\kern -0.18em \PB}{}}}\xspace}

\def\BorBbar    {\kern 0.18em\optbar{\kern -0.18em B}{}\xspace}

  \def\Y#1S{\ensuremath{\PUpsilon{(#1S)}}\xspace}

\def\Lbar        {{\ensuremath{\kern 0.1em\overline{\kern -0.1em\PLambda}}}\xspace}
\def\LorLbar    {\kern 0.18em\optbar{\kern -0.18em \PLambda}{}\xspace}

\def\to                 {\ensuremath{\rightarrow}\xspace}

\def\AT#1     {\ensuremath{A_{\mathrm{T}}^{#1}}\xspace}           

\def\C#1      {\ensuremath{\mathcal{C}_{#1}}\xspace}                       
\def\Cp#1     {\ensuremath{\mathcal{C}_{#1}^{'}}\xspace}                    
\def\Ceff#1   {\ensuremath{\mathcal{C}_{#1}^{\mathrm{(eff)}}}\xspace}        
\def\Cpeff#1  {\ensuremath{\mathcal{C}_{#1}^{'\mathrm{(eff)}}}\xspace}       
\def\Ope#1    {\ensuremath{\mathcal{O}_{#1}}\xspace}                       
\def\Opep#1   {\ensuremath{\mathcal{O}_{#1}^{'}}\xspace}                    

\newcommand{\tev}{\ifthenelse{\boolean{inbibliography}}{\ensuremath{~T\kern -0.05em eV}\xspace}{\ensuremath{\mathrm{\,Te\kern -0.1em V}}}\xspace}
\newcommand{\gev}{\ensuremath{\mathrm{\,Ge\kern -0.1em V}}\xspace}
\newcommand{\mev}{\ensuremath{\mathrm{\,Me\kern -0.1em V}}\xspace}
\newcommand{\kev}{\ensuremath{\mathrm{\,ke\kern -0.1em V}}\xspace}
\newcommand{\ev}{\ensuremath{\mathrm{\,e\kern -0.1em V}}\xspace}
\newcommand{\gevc}{\ensuremath{{\mathrm{\,Ge\kern -0.1em V\!/}c}}\xspace}
\newcommand{\mevc}{\ensuremath{{\mathrm{\,Me\kern -0.1em V\!/}c}}\xspace}
\newcommand{\gevcc}{\ensuremath{{\mathrm{\,Ge\kern -0.1em V\!/}c^2}}\xspace}
\newcommand{\gevgevcccc}{\ensuremath{{\mathrm{\,Ge\kern -0.1em V^2\!/}c^4}}\xspace}
\newcommand{\mevcc}{\ensuremath{{\mathrm{\,Me\kern -0.1em V\!/}c^2}}\xspace}

\def\gsim{{~\raise.15em\hbox{$>$}\kern-.85em
          \lower.35em\hbox{$\sim$}~}\xspace}
\def\lsim{{~\raise.15em\hbox{$<$}\kern-.85em
          \lower.35em\hbox{$\sim$}~}\xspace}

\newcommand{\Real}{\ensuremath{\mathcal{R}e}\xspace}
\newcommand{\Imag}{\ensuremath{\mathcal{I}m}\xspace}

\def\fortran    {\mbox{\textsc{Fortran}}\xspace}

\def\tell1  {TELL1\xspace}
\def\ukl1   {UKL1\xspace}

\preprintnumber{CERN-TH-2017-167, NORDITA 2017-077}

\title{FlavBit: A GAMBIT module for computing flavour observables and likelihoods}

\author{The GAMBIT Flavour Workgroup:
Florian U.~Bernlochner\thanksref{inst:z} \and
Marcin Chrz\k{a}szcz\thanksref{inst:a,inst:b,e1} \and
Lars A.~Dal\thanksref{inst:c} \and
Ben Farmer\thanksref{inst:d,inst:e} \and
Paul Jackson\thanksref{inst:f,inst:g} \and
Anders Kvellestad\thanksref{inst:h} \and
Farvah Mahmoudi\thanksref{inst:i,inst:j,e2,e5} \and
Antje Putze\thanksref{inst:l} \and
Christopher Rogan\thanksref{inst:m} \and
Pat Scott\thanksref{inst:n,e3} \and
Nicola Serra\thanksref{inst:a,e4} \and
Christoph Weniger\thanksref{inst:o} \and
Martin White\thanksref{inst:f,inst:g}
}

\institute{%
  \bonn\label{inst:z} \and
  \zurich\label{inst:a} \and
  \krakow\label{inst:b} \and
  \oslo\label{inst:c} \and
  \okc\label{inst:d} \and
  \su\label{inst:e} \and
  \adelaide\label{inst:f} \and
  \coepp\label{inst:g} \and
  \nordita\label{inst:h} \and
  \lyon\label{inst:i} \and
  \cernth\label{inst:j} \and
  \annecy\label{inst:l} \and
  \harvard\label{inst:m} \and
  \imperial\label{inst:n} \and
  \grappa\label{inst:o}
}

\thankstext{e1}{mchrzasz@cern.ch}
\thankstext{e2}{nazila@cern.ch}
\thankstext{e3}{p.scott@imperial.ac.uk}
\thankstext{e4}{nicola.serra@cern.ch}
\thankstext[*]{e5}{Also \iuf.}

\titlerunning{FlavBit}
\authorrunning{GAMBIT Flavour Workgroup}

\date{Received: date / Accepted: date}

\begin{document}

\maketitle

\begin{abstract}
Flavour physics observables are excellent probes of new physics up to very high energy scales.
Here we present \flavbit, the dedicated flavour physics module of the global-fitting package \gambit.
\flavbit includes custom implementations of various likelihood routines for a wide range of flavour observables, including detailed uncertainties and correlations associated with LHCb measurements of rare, leptonic and semileptonic decays of $B$ and $D$ mesons, kaons and pions.  It provides a generalised interface to external theory codes such as \superiso, allowing users to calculate flavour observables in and beyond the Standard Model, and then test them in detail against all relevant experimental data. We describe \flavbit and its constituent physics in some detail, then give examples from supersymmetry and effective field theory illustrating how it can be used both as a standalone library for flavour physics, and within \GB.

\end{abstract}

\tableofcontents

\section{Introduction}
\label{sec:introduction}

Precise measurement of flavour observables is a powerful indirect probe of physics beyond the Standard Model (SM), as new heavy particles predicted by extensions of the SM can contribute to the amplitudes of observables as virtual particles.  Flavour observables are therefore sensitive to much higher energy scales than direct searches for new particles. Moreover, rare decays, such as Flavour Changing Neutral Currents (FCNCs), are loop suppressed in the SM. As a consequence, the SM decay rates are small, and could be comparable in magnitude to contributions from new heavy states, allowing stringent constraints to be placed on the parameters of theories for new physics.
It is therefore crucial to consider constraints from flavour physics when studying scenarios beyond the SM. The correlations between the different flavour observables, and the interplay between flavour measurements and direct searches at collider experiments, are key tools in the search for new physics, and its eventual understanding.

Public packages exist for carrying out SM and BSM flavour fits in terms of Wilson coefficients \cite{Serra:2016ivr,flavio,hepfit}, but so far no general package exists for both computing Wilson coefficients and carrying out a global fit.  In this article we present \flavbit, a flavour physics library designed in the context of the Global And Modular BSM Inference Tool (\gambit) framework~\cite{gambit}, but also usable in standalone form. \flavbit allows users to predict flavour physics observables in various models, using external programs such as \superiso~\cite{Mahmoudi:2007vz,Mahmoudi:2008tp,Mahmoudi:2009zz}, and then calculate combined likelihoods for arbitrary combinations of the observables. \flavbit takes into account all theoretical and experimental correlations between the different observables.  The resulting likelihoods can be incorporated into the \GB global likelihood to scan the parameter spaces of various models for new physics \cite{gambit,ScannerBit,SSDM,CMSSM,MSSM}, taking into account complementary constraints from direct production \cite{ColliderBit}, dark matter searches \cite{DarkBit}, and SM and related precision measurements \cite{SDPBit}.

Recently, some measurements of flavour observables, mainly from LHCb~\cite{Aaij:2015esa,Aaij:2015yra,Aaij:2014ora,Aaij:2013qta} and $B$ factories~\cite{Lees:2012xj,Lees:2013uzd,Huschle:2015rga,Abdesselam:2016cgx,Abdesselam:2016llu}, have shown tension with their predicted values in the SM.  It is still unclear if these might be accommodated in the SM by larger-than-expected QCD effects, statistical fluctuations or some combination thereof. Nonetheless, these tensions certainly provide motivation for continued interest and effort in careful combination and cross-correlation of flavour observables with each other, and with searches for new physics in other sectors.  We include these measurements in \flavbit.

This paper is organised as follows. In Sec.~\ref{sec:theory} we provide the general theoretical background of the scheme by which we compute flavour observables, before providing a brief synopsis in Sec.~\ref{sec:framework} of the broader global-fitting framework within which \flavbit sits. In Sec.~\ref{sec:obs} we discuss the predictions and measurements of individual observables included in \flavbit \textsf{1.0.0}, and highlight aspects of new physics models to which the different measurements are sensitive. Sec.~\ref{sec:like} gives details of the likelihood calculations that \flavbit performs. Sec.~\ref{sec:examples} gives some usage examples, both in standalone mode and with \gambit proper.  Sec.~\ref{sec:summary} summarises our conclusions, and Appendix \ref{glossary} gives a glossary of relevant \GB terminology helpful for reading this paper.

The \flavbit source code is freely available from \href{http://gambit.hepforge.org}{gambit.hepforge.org} under the terms of the standard 3-clause BSD license.\footnote{\href{http://opensource.org/licenses/BSD-3-Clause}{http://opensource.org/licenses/BSD-3-Clause}.  Note that \textsf{fjcore} \cite{Cacciari:2011ma} and some outputs of \flexiblesusy \cite{Athron:2014yba} (incorporating routines from \SOFTSUSY \cite{Allanach:2001kg}) are also shipped with \GB \textsf{1.0}.  These code snippets are distributed under the GNU General Public License (GPL; \href{http://opensource.org/licenses/GPL-3.0}{http://opensource.org/licenses/GPL-3.0}), with the special exception, granted to \GB by the authors, that they do not require the rest of \GB to inherit the GPL.}

\section{Theoretical framework}
\label{sec:theory}

Our theoretical framework for studying rare decay observables is based on the effective Hamiltonian approach, which provides a simple formulation that can be easily extended to incorporate contributions from new physics. In this formulation, the low- and high-energy effects are separated using the Operator Product Expansion method.  Cross-sections for transitions from initial states $i$ to final states $f$ are proportional to squared matrix elements $|\langle f |{\cal H}_{\rm eff}|i\rangle|^2$, where the effective Hamiltonian ${\cal H}_{\rm eff}$ for $b \to s$ transitions is given by
\begin{equation}
\mathcal{H}_{\rm eff}  =  -\frac{4G_{F}}{\sqrt{2}} V_{tb} V_{ts}^{*} \sum_{i=1}^{10} \Bigl(C_{i}(\mu) \mathcal{O}_i(\mu)+C'_{i}(\mu) \mathcal{O}'_i(\mu)\Bigr)\;.
\end{equation}
Here $G_\text{F}$ is the Fermi constant, $\mu$ is the energy scale at which calculations are to be performed, and $V_{tb}$ and $V_{ts}$ are the usual CKM matrix elements.  The $C_i$ are Wilson coefficients, which incorporate the influence of small-scale physics due to heavy states that have been integrated out in the effective theory; their values can be calculated using perturbative methods.  The $\mathcal{O}_i$ are local operators representing long-distance interactions. The most relevant operators for the FCNC rare $B$ decays are
\begin{align}
\label{physical_basis}
\mathcal{O}_1& =  (\bar{s} \gamma_{\mu} T^a P_L c) (\bar{c} \gamma^{\mu} T^a P_L b)\;,\nonumber\\
\mathcal{O}_2& = (\bar{s} \gamma_{\mu} P_L c) (\bar{c} \gamma^{\mu} P_L b)\;,  \nonumber\\
\mathcal{O}_3& =  (\bar{s} \gamma_{\mu} P_L b) \sum_q (\bar{q} \gamma^{\mu} q)\;, \nonumber\\
\mathcal{O}_4& = (\bar{s} \gamma_{\mu} T^a P_L b) \sum_q (\bar{q} \gamma^{\mu} T^a q)\;,\nonumber\\
\mathcal{O}_5& =  (\bar{s} \gamma_{\mu_1} \gamma_{\mu_2} \gamma_{\mu_3} P_L b)
                  \sum_q (\bar{q} \gamma^{\mu_1} \gamma^{\mu_2} \gamma^{\mu_3} q)\;, \nonumber\\
\mathcal{O}_6& = (\bar{s} \gamma_{\mu_1} \gamma_{\mu_2} \gamma_{\mu_3} T^a P_L b)
                  \sum_q (\bar{q} \gamma^{\mu_1} \gamma^{\mu_2} \gamma^{\mu_3} T^a q)\;,\nonumber\\
\mathcal{O}_7& = \frac{e}{(4\pi)^2} m_b (\overline{s} \sigma^{\mu\nu} P_R b) F_{\mu\nu} \;, \nonumber\\
\mathcal{O}_8& = \frac{g}{(4\pi)^2} m_b (\bar{s} \sigma^{\mu \nu} T^a P_R b) G_{\mu \nu}^a \;,  \nonumber\\
\mathcal{O}_9& =  \frac{e^2}{(4\pi)^2} (\overline{s} \gamma^\mu P_L b) (\bar{\ell} \gamma_\mu \ell) \;,   \nonumber\\
\mathcal{O}_{10}& =  \frac{e^2}{(4\pi)^2} (\overline{s} \gamma^\mu P_L b) (\bar{\ell} \gamma_\mu \gamma_5 \ell) \;,
\end{align}
where the sums run over $q=u, d, s, c, b$, $m_b$ denotes the $b$ quark mass, $T^a$ are the SU(3)$_c$ generators, $F_{\mu\nu}$ and $G_{\mu \nu}^a$ are the photon and gluon stress-energy tensors respectively, and $g$ is the strong coupling. A similar set of operators can also be defined for $b \to d$ transitions.

This formalism can be easily extended to incorporate effects of new physics, through additional contributions to the Wilson coefficients or the introduction of additional long-distance operators. For instance, the primed versions of these operators are chirality-flipped compared to the non-primed ones, and are highly suppressed in the SM. The scalar ($\mathcal{Q}_{1}$) and pseudoscalar ($\mathcal{Q}_{2}$) operators
\begin{align}
\label{physical_basis_extra}
\mathcal{Q}_{1}& = \frac{e^2}{(4\pi)^2}(\bar{s} P_R b)(\bar{\ell}\,\ell) \;, \\
\mathcal{Q}_{2}& = \frac{e^2}{(4\pi)^2}(\bar{s} P_R b)(\bar{\ell}\gamma_5 \ell) \;,
\end{align}
are absent in the SM, but receive large contributions in many models with an extended Higgs sector.

The Wilson coefficients are calculated by requiring matching between the high-scale theory and the low-energy effective theory at the scale $\mu_W$, which is of the order of the $W$ mass. Using the renormalisation group equations of the effective theory, they are then evolved to the scale $\mu_b$ (of the order of the $b$ quark mass), which is the relevant scale for $B$ physics calculations.

In order to compute the matrix element $\langle f| {\cal H}_{\rm eff}|i\rangle$, which describes the transition from the initial state $|i\rangle$ to the final state $|f\rangle$, in addition to the relevant Wilson coefficients $C_i$, we need to evaluate the hadronic matrix elements $\langle f| {\cal O}_i |i\rangle$, which are usually the main source of uncertainties. These elements lead to decay constants and form factors that must be computed with techniques from non-perturbative QCD.

\section{Computational framework}
\label{sec:framework}

The \gambit framework defines two sorts of functions that can be used to calculate physical observables or other quantities required for computing them:
\begin{description}
\item[\doublecross{module functions}{module function}:] functions written in \Cpp and contained within a \gambit \cross{module}.
\item[\doublecross{backend functions}{backend function}:] external library functions provided by a \cross{backend}, such as \superiso or \FH.
\end{description}
For ease of reference, here we \doublecross{highlight}{glossary} and link specific \GB terms to their entries in the glossary, found in Appendix \ref{glossary}.

When writing \gambit module functions, the author assigns each a \cross{capability}, which describes what the function can calculate.  This may be an observable, e.g.\ a particular branching fraction for a given rare $B$ decay, or a likelihood, e.g.\ the combined likelihood defined using a set of rare decays. Module functions can be declared to have \doublecross{dependencies}{dependency} on the results of other module functions, which they indicate by specifying the \cross{capability} of the module function that must be used to fill the dependency.  Dependencies may be filled by any function within \GB that has the requisite capability, whether or not it is part of the same \GB \cross{module} as the dependent function.  Module functions may also have \doublecross{backend requirements}{backend requirement}, which are satisfied by functions from \cross{backend} libraries.  For example, in \flavbit \textsf{1.0.0}, \superiso supplies many of the backend requirements of the module functions that calculate observables.

\flavbit notifies \gambit of its available module functions and their capabilities, dependencies and backend requirements. The user tells \GB that they want to compute a given set of observables and likelihoods in a given scan, and the \GB Core identifies the necessary module functions and runs its \cross{dependency resolution} routines.  These hook the module functions up to each other and run them in an order that ensures that all dependencies are computed before the functions that depend on them. Full details of this process can be found in the main \gambit paper~\cite{gambit}.

In standalone mode, users can just call the module functions of \flavbit directly, providing any required dependencies and backend requirements manually.

\section{Observables}
\label{sec:obs}

In this section we discuss the observables included in \flavbit and their relevance for searches for new physics.

The most important observables are the rare decays $B \to X_s \gamma$, $B^0_s\to\mu^+\mu^-$ and $B^0\to K^{*0}\mu^+\mu^-$, as well as tree level decays such as $B^\pm \to \tau \nu_\tau$ and $B \to D^{(*)} \ell\nu_\ell$.\footnote{Here $D^{(*)}$, $B^\pm$ and $\ell$ are shorthand notations. The first indicates that we are referring to both $B \to D \ell\nu_\ell$ and $B \to D^* \ell\nu_\ell$, but as distinct processes. The same is true of the second notation, which indicates that we are referring to both the original process and its CP conjugate, distinctly. In contrast, when referring to specific rates, $\ell$ is typically used to indicate that the final state does not distinguish between $\ell = e$ and $\ell=\mu$.  Some groups use this notation to refer to a sum over all final states involving electrons and muons, others use it to refer to the average.  The PDG uses the former notation, which we follow in this paper except where explicitly noted otherwise.}

Here we discuss the calculation of the different observables in four groups: tree-level leptonic and semi-leptonic decays (Sec.\ \ref{sec:obs_tree}), electroweak penguin transitions (Sec.\ \ref{sec:obs_penguin}), rare purely leptonic decays (Sec.\ \ref{sec:obs_rare}), and other flavour observables (Sec.\ \ref{sec:obs_other}). In these sections we outline the calculations required to predict each observable from theory; further details can be found in Ref.~\cite{Mahmoudi:2008tp}. While for simplicity we present only the leading order expressions in this paper, in \flavbit itself we use the full calculations at the highest available accuracy.

The tree-level category includes $B$ and $D$ decays to leptons with an accompanying hadron and/or a neutrino in the final state.  Observables in this category are the branching fractions for processes such as $B^\pm \to \tau \nu_\tau$, $B \to D^{(*)} \tau \nu_\tau$ and $B \to D^{(*)} \ell \nu_\ell$.  The electroweak penguin category includes the rare decays $B \to M \ell^+\ell^-$ (with $M$ another meson lighter than the $B$), in particular the angular observables of the decay $B^0 \to K^{*0} \mu^+\mu^-$. The rare fully-leptonic category includes $B$ decays with only leptons in the final state, such as $B^0_{(s)} \to \mu^+ \mu^-$.  The fourth and final category includes $b\to s$ transitions in the radiative decays $B \to X_s \gamma$, the mass difference between the heavy $B_H$ and light $B_L$ eigenstates of the $B^0_s$ system ($\Delta M_s$), and decays of kaons and pions, in particular the leptonic decay ratio ${\cal B}(K^\pm\to \mu \nu_\mu)/{\cal B}(\pi^\pm\to \mu \nu_\mu)$.  Note that \flavbit does not incorporate the anomalous magnetic moment of the muon, as this is dealt with in \precisionbit \cite{SDPBit}.

\subsection{Interfaces to external codes}

Theoretical predictions of observables in \flavbit are predominantly obtained through interfaces to external codes. Some predictions of flavour observables are available from \FH \cite{Heinemeyer:1998yj}, for the SM and minimal supersymmetric SM (MSSM).\footnote{The \GB interface to \FH is described in detail in Sec 3.1.3 of Ref.\ \cite{SDPBit}.}  In \flavbit \textsf{1.0.0}, most observable calculations refer to \superiso \textsf{3.6} \cite{Mahmoudi:2007vz,Mahmoudi:2008tp,Mahmoudi:2009zz}.

The interface to \superiso operates via the function \cpp{SI_fill} (see Table \ref{tab:flavourobswoWC}), which provides the \cpp{SuperIso\_modelinfo}.  This function fills a \superiso\ \cpp{parameters} structure, which is passed back to various other \superiso functions to compute observables.  Observables that are calculated directly from the input model parameters (Table \ref{tab:flavourobswoWC}) are distinguished from those that involve the calculation of intermediate Wilson coefficients (Tables \ref{tab:flavourobsWC} and \ref{tab:flavourobsBKstar}).  In \flavbit \textsf{1.0.0}, observables are implemented for MSSM models (`\textsf{MSSM63atQ}' and descendants; see \cite{gambit}), and for a flavour EFT model (`\textsf{WC}') where the Wilson coefficients are specified directly as model parameters, and scanned over.

The design of \flavbit and its interface to \superiso make extending \flavbit to other models quite straightforward, either by computing Wilson coefficients `upstream' from fundamental parameters, or by constructing the \cpp{SuperIso\_modelinfo} to fit the model under investigation.  \cpp{SI_fill} deals with the majority of the model-dependence in each calculation, importing different masses and couplings from \cpp{SpecBit} depending on the model being scanned, and using them to set various flags and member variables of the \cpp{SuperIso\_modelinfo}.

\cpp{SI_fill} has a single option configurable from the master \YAML file of a given scan: a boolean flag \yaml{take_b_pole_mass_from_spectrum}.  This option allows the user to choose between \superiso's internal calculation of the $b$ quark pole mass (based on the \MSbar mass imported from \GB), or \GB's own $b$ pole mass calculation provided by \specbit \cite{SDPBit}.  Depending on the spectrum generator chosen in \specbit, the standard 2-loop conversion from \MSbar to pole mass included in \superiso may be a more accurate choice for precision $B$ physics than other calculations, even if the other calculation includes higher-order corrections.  This is because the $b$ pole is sufficiently close to the QCD scale that problems with the perturbative expansion required to compute it start to show already at 3 loops \cite{Olive:2016xmw}, such that the formal error on the $b$ pole mass associated with truncating the asymptotic series may already be larger when truncating at 3 rather than 2 loops.  This means that although 3-loop QCD RGEs remain preferable, 2-loop self energies give a more precise value for the $b$ pole, and should be preferred for $B$ physics calculations.  In \flavbit \textsf{1.0.0}, \yaml{take_b_pole_mass_from_spectrum} therefore defaults to \yaml{false}.\footnote{Note that \superiso only actually uses the $b$ pole mass for computing the 1S mass, which is better-behaved than the pole mass and preferable for observable calculations.}

\begin{table*}[tp]
\centering
\small{
\begin{tabular}{l|p{7.1cm}|l|l}
  \textbf{Capability}
      & \multirow{2}{*}{\parbox{7.1cm}{\textbf{Function (Return Type)}: \\ \textbf{Brief Description}}} & \textbf{Dependencies (Model)}
      & \multirow{2}{*}{\parbox{2cm}{\textbf{Backend\\ requirements}}}
      \\ & & &
  \\ \hline
      \cpp{SuperIso\_modelinfo}
      & \multirow{2}{*}{\parbox{7.1cm}{\cpp{SI\_fill} (\cpp{parameters}):
      \\ Fills the \superiso structure. Key routine of the \superiso interface.}}
      & \cpp{MSSM\_spectrum} {\scriptsize(\textsf{MSSM63atQ})} & \cpp{Init\_param}
      \\ & & \cpp{SM\_spectrum} {\scriptsize(\textsf{WC})} & \cpp{slha\_adjust}
      \\ & & \cpp{W_plus_decay_rates} & \cpp{mb_1S}
      \\ & & \cpp{Z_decay_rates} &
  \\ \hline
      \cpp{Dstaunu}
      & \multirow{2}{*}{\parbox{7.1cm}{\cpp{SI\_Dstaunu} (\cpp{double}):
              \\ Computes the branching fraction of $D^\pm_s\to \tau \nu_\tau$.}}
          & \cpp{SuperIso\_modelinfo} & \cpp{Dstaunu}
          \\ & &
    \\ \hline
      \cpp{Dsmunu}
      & \multirow{2}{*}{\parbox{7.1cm}{\cpp{SI\_Dsmunu} (\cpp{double}):
              \\ Computes the branching fraction of $D^\pm_s\to \mu \nu_\mu$.}}
          & \cpp{SuperIso\_modelinfo} & \cpp{Dsmunu}
          \\ & &
    \\ \hline
      \cpp{Dmunu}
      & \multirow{2}{*}{\parbox{7.1cm}{\cpp{SI\_Dmunu} (\cpp{double}):
              \\ Computes the branching fraction of $D^\pm\to \mu \nu_\mu$.}}
          & \cpp{SuperIso\_modelinfo} & \cpp{Dmunu}
          \\ & &
    \\ \hline
    \cpp{Btaunu}
      & \multirow{2}{*}{\parbox{7.1cm}{\cpp{SI\_Btaunu} (\cpp{double}):
              \\ Computes the branching fraction of $B^\pm\to \tau \nu_\tau$.}}
          & \cpp{SuperIso\_modelinfo} & \cpp{Btaunu}
          \\ & &
  \\ \hline
     \cpp{BDtaunu}
      & \multirow{2}{*}{\parbox{7.1cm}{\cpp{SI\_BDtaunu} (\cpp{double}):
              \\ Computes the branching fraction of $B\to D \tau \nu_\tau$.}}
          & \cpp{SuperIso\_modelinfo} & \cpp{BRBDlnu}
          \\ & &
  \\ \hline
     \cpp{BDmunu}
      & \multirow{2}{*}{\parbox{7.1cm}{\cpp{SI\_BDmunu} (\cpp{double}):
              \\ Computes the branching fraction of $B\to D \mu \nu_\mu$.}}
          & \cpp{SuperIso\_modelinfo} & \cpp{BRBDlnu}
          \\ & &
  \\ \hline
     \cpp{BDstartaunu}
      & \multirow{2}{*}{\parbox{7.1cm}{\cpp{SI\_BDstartaunu} (\cpp{double}):
              \\ Computes the branching fraction of $B\to D^* \tau \nu_\tau$.}}
          & \cpp{SuperIso\_modelinfo} & \cpp{BRBDstarlnu}
          \\ & &
  \\ \hline
     \cpp{BDstarmunu}
      & \multirow{2}{*}{\parbox{7.1cm}{\cpp{SI\_BDstarmunu} (\cpp{double}):
              \\ Computes the branching fraction of $B\to D^* \mu \nu_\mu$.}}
          & \cpp{SuperIso\_modelinfo} & \cpp{BRBDstarlnu}
          \\ & &
  \\ \hline
     \cpp{RD}
      & \multirow{2}{*}{\parbox{7.1cm}{\cpp{SI\_RD} (\cpp{double}):
              \\ Computes the ratio $\mathcal{B}(B\to D \tau \nu_\tau)/\mathcal{B}(B\to D l \nu_l)$, where $\ell = \mu$ or $e$ and the result is the same for each.}}
          & \cpp{SuperIso\_modelinfo} & \cpp{BDtaunu\_BDenu}
          \\ & &
          \\ & &
  \\ \hline
     \cpp{RDstar}
      & \multirow{2}{*}{\parbox{7.1cm}{\cpp{SI\_RDstar} (\cpp{double}):
              \\ Computes the ratio $\mathcal{B}(B\to D^* \tau \nu_\tau)/\mathcal{B}(B\to D^* l \nu_l)$, where $\ell = \mu$ or $e$ and the result is the same for each.}}
          & \cpp{SuperIso\_modelinfo} & \cpp{BDstartaunu\_}
          \\ & & & \ \cpp{BDstarenu}
          \\ & & &
  \\ \hline
      \cpp{Rmu}
      & \multirow{2}{*}{\parbox{7.1cm}{\cpp{SI\_Rmu} (\cpp{double}):
              \\ Computes the ratio $\mathcal{B}(K^\pm\to \mu \nu_\mu)/\mathcal{B}(\pi^\pm\to \mu \nu_\mu)$.}}
          & \cpp{SuperIso\_modelinfo} & \cpp{Kmunu\_pimunu}
          \\ & &
   \\ \hline
      \cpp{Rmu23}
      & \multirow{2}{*}{\parbox{7.1cm}{\cpp{SI\_Rmu23} (\cpp{double}):
              \\ Computes the observable $R_{\mu23}$ (Eq.\ \ref{eq:rmu23}).}}
          & \cpp{SuperIso\_modelinfo} & \cpp{Rmu23}
          \\ & &
    \\ \hline
      \cpp{FH\_FlavourObs}
      & \multirow{2}{*}{\parbox{7.1cm}{\cpp{FH\_FlavourObs} (\cpp{fh\_FlavourObs}):
              \\ Computes the \FH flavour observables.}}
          & & \cpp{FHFlavour}
          \\ & &
          \\ \hline
      \cpp{deltaMs}
      & \multirow{3}{*}{\parbox{7.1cm}{\cpp{FH\_DeltaMs} (\cpp{double}):
              \\ Extracts the \FH MSSM prediction for the $B_s$--$\bar{B}_s$ mass difference $\Delta M_s$ (in ps$^{-1}$).}}
          & \cpp{FH\_FlavourObs} &
          \\ & &
          \\ & &
          \\ \hline
\end{tabular}
}
\caption{Observable capabilities of \flavbit that do not involve Wilson coefficients. Details of the \cpp{fh\_FlavourObs} structure can be found in Table \ref{tab:fhFlavourObs}. \label{tab:flavourobswoWC}}
\end{table*}

\subsection{Tree-level leptonic and semi-leptonic decays}
\label{sec:obs_tree}

Decays of $B$ mesons with leptons and neutrinos in the final state proceed via tree-level charged currents. They have been intensively studied at $B$ factories (Babar, Belle and CLEO) for the determination of the elements $V_{cb}$ and $V_{ub}$ of the CKM matrix.

The rate of the semi-leptonic decay $B \to M \ell \nu_{\ell}$ in the SM is
\begin{equation}
\label{eq:bxlnu}
\frac{d\Gamma}{dq^2}=\frac{G_F^2 |V_{qb}^2|}{192\pi^3m_B^3}{\cal K}(m^2_B, m^2_M, q^2) {\cal F}^{(2)}(q^2)\;,
\end{equation}
where $q^\mu=p_B^\mu-p^\mu_M$ is the momentum transfer, $V_{qb}$ is the CKM element corresponding to the flavour of $M$, ${\cal K}$ is a phase-space factor and ${\cal F}^{(2)}(q^2)$ is a combination of
form factors~\cite{Sakaki:2013bfa}.

These decays are sensitive to charged-current contributions from new particles. For example, the charged Higgs in the two Higgs doublet model (2HDM) (see e.g.  Refs.~\cite{Crivellin:2015hha,Freytsis:2015qca,Buras:2013ooa,Eberhardt:2013uba}), right-handed currents via the contribution of the charged mediator $W_R$~\cite{Das:2016vkr,Buras:2013ooa}, new left-handed heavy bosons $W^{\prime}$~\cite{Greljo:2015mma,Boucenna:2016qad} and leptoquarks (see e.g. Refs.~\cite{Becirevic:2016oho,Dorsner:2016wpm}) can also modify the value of this observable.

The decays $B^\pm\to \ell \nu_{\ell}$ also proceed via tree-level charged currents. The branching fraction is
\begin{align}
{\cal B}(B^+\to \ell^+\nu_{\ell}) = \frac{G_F^2 m_Bm^2_{\ell}}{8\pi}\left( 1-\frac{m_{\ell}^2}{m_B^2}\right)^2 f_{B}^2 |V_{ub}|^2 \tau_{B},
\label{eq:Blnu}
\end{align}
where $f_B$ is the meson decay constant and $\tau_B$ is the lifetime of the $B^+$. This decay is sensitive to the CKM element $V_{ub}$. The charged Higgs sector of the 2HDM can again provide substantial contributions, as can new charged gauge bosons like the $W^{\prime}$ and $W_R$ of the left-right symmetric model~\cite{Bona:2009cj}.
Compared to the case where $\ell=\tau$, the decays with $\ell=e$ and $\ell=\mu$ have much smaller branching fractions, as they are helicity-suppressed. For this reason, at present only upper limits are available for the decays to light leptons. Although we provide routines to predict the values of all three in \flavbit, we only incorporate the tauonic version into the resulting likelihood.

Similarly, the decays $D^\pm_{(s)}\to \ell \nu_\ell$ are mediated by the $W$ boson in the SM. The branching fractions can be obtained from Eq.~\ref{eq:Blnu} after the replacement $B\to D_{(s)}$ and swapping in the relevant CKM element. These decays have been traditionally used to measure the $D_{(s)}$ meson decay constant. However, the charged Higgs boson in the 2HDM would also mediate these decays, so they can provide complementary constraints to the analogous $B$ meson decay~\cite{Akeroyd:2009tn}.

As shown in Table \ref{tab:flavourobswoWC}, \flavbit provides functions capable of computing branching fractions for $D^\pm_s\to \tau \nu_\tau$ (\cpp{Dstaunu}), $D^\pm_s\to \mu \nu_\mu$ (\cpp{Dsmunu}), $D^\pm\to \mu \nu_\mu$ (\cpp{Dmunu}), $B^\pm\to \tau \nu_\tau$ (\cpp{Btaunu}), $B\to D \tau \nu_\tau$ (\cpp{BDtaunu}), $B\to D \mu \nu_\mu$ (\cpp{BDmunu}), $B\to D^* \tau \nu_\tau$ (\cpp{BDstartaunu}) and $B\to D^* \mu \nu_\mu$ (\cpp{BDstarmunu}). It can also compute $R_{D^{(*)}} \equiv \mathcal{B}(B\to D^{(*)} \tau \nu_\tau)/\mathcal{B}(B\to D^{(*)} l \nu_l)$, designated by capabilities \cpp{RD} and \cpp{RDstar}. Here $\ell$ in $R_{D^{(*)}}$ refers to either $\mu$ or $e$, not their sum (the branching fractions $B\to D^{(*)} l \nu_l$ are identical for $e$ and $\mu$, as both are effectively massless in the $B$ system).

\begin{table*}[tp]
\centering
\small{
\begin{tabular}{l|p{6.4cm}|l|l}
  \textbf{Capability}
      & \multirow{2}{*}{\parbox{6.4cm}{\textbf{Function (Return Type)}: \\ \textbf{Brief Description}}} & \textbf{Dependencies}
      & \multirow{2}{*}{\parbox{2cm}{\textbf{Backend\\ requirements}}}
      \\ & & &
  \\ \hline
  \cpp{bsgamma}
      & \multirow{2}{*}{\parbox{6.4cm}{\cpp{SI\_bsgamma} (\cpp{double}):
              \\ Computes the inclusive branching fraction of $B\to X_s \gamma$ for $E_\gamma>1.6$\,GeV.}}
          & \cpp{SuperIso\_modelinfo} & \cpp{bsgamma_CONV}
          \\ & & &
          \\ & & &
      \\\cmidrule{2-4}
      & \multirow{2}{*}{\parbox{6.4cm}{\cpp{FH\_bsgamma} (\cpp{double}):
              \\ Extracts the total inclusive branching fraction of $B \to X_s\gamma$ in the MSSM from \FH.}}
          & \cpp{FH_FlavourObs} &
          \\ & & &
          \\ & & &
  \\ \hline
     \cpp{delta0}
      & \multirow{2}{*}{\parbox{6.4cm}{\cpp{SI\_delta0} (\cpp{double}):
              \\ Computes the isospin asymmetry of $B\to K^* \gamma$.}}
          & \cpp{SuperIso\_modelinfo} & \cpp{delta0\_CONV}
          \\ & & &

  \\ \hline
       \cpp{Bsmumu\_untag}
      & \multirow{2}{*}{\parbox{6.4cm}{\cpp{SI\_Bsmumu\_untag} (\cpp{double}):
              \\ Computes the $CP$-averaged branching fraction of $B^0_s\to \mu^+ \mu^-$.}}
          & \cpp{SuperIso\_modelinfo} & \cpp{Bsll_untag_CONV}
          \\ & & &
          \\ & & &
      \\\cmidrule{2-4}
      & \multirow{2}{*}{\parbox{6.4cm}{\cpp{FH\_Bsmumu} (\cpp{double}):
              \\ Extracts the $CP$-averaged branching fraction of $B^0_s\to \mu^+ \mu^-$ in the MSSM from \FH.}}
          & \cpp{FH_FlavourObs} &
          \\ & & &
          \\ & & &
  \\ \hline
       \cpp{Bsee\_untag}
      & \multirow{2}{*}{\parbox{6.4cm}{\cpp{SI\_Bsee\_untag} (\cpp{double}):
              \\ Computes the $CP$-averaged branching fraction of $B^0_s\to e^+ e^-$.}}
          & \cpp{SuperIso\_modelinfo} & \cpp{Bsll_untag_CONV}
          \\ & & &
          \\ & & &
  \\ \hline
        \cpp{Bmumu}
      & \multirow{2}{*}{\parbox{6.4cm}{\cpp{SI\_Bmumu} (\cpp{double}):
              \\ Computes the branching fraction of $B^0\to \mu^+ \mu^-$.}}
          & \cpp{SuperIso\_modelinfo} & \cpp{Bll_CONV}
          \\ & & &
  \\ \hline
        \cpp{BRBXsmumu\_lowq2}
      & \multirow{2}{*}{\parbox{6.4cm}{\cpp{SI\_BRBXsmumu\_lowq2} (\cpp{double}):
              \\ Computes the inclusive low-$q^2$ branching fraction of $B\to X_s\mu^+ \mu^-$.}}
          & \cpp{SuperIso\_modelinfo} &  \cpp{BRBXsmumu\_lowq2\_CONV}
          \\ & & &
          \\ & & &
  \\ \hline
        \cpp{BRBXsmumu\_highq2}
      & \multirow{2}{*}{\parbox{6.4cm}{\cpp{SI\_BRBXsmumu\_highq2} (\cpp{double}):
              \\ Computes the inclusive high-$q^2$ branching fraction of $B\to X_s\mu^+ \mu^-$.}}
          & \cpp{SuperIso\_modelinfo} & \cpp{BRBXsmumu\_high2\_CONV}
          \\ & & &
          \\ & & &  \\ \hline
        \cpp{A\_BXsmumu\_lowq2}
      & \multirow{2}{*}{\parbox{6.4cm}{\cpp{SI\_A\_BXsmumu\_lowq2} (\cpp{double}):
              \\ Computes the low-$q^2$ forward-backward asymmetry of $B\to X_s\mu^+ \mu^-$.}}
          & \cpp{SuperIso\_modelinfo} &  \cpp{A\_BXsmumu\_lowq2\_CONV}
          \\ & & &
          \\ & & &  \\ \hline
         \cpp{A\_BXsmumu\_highq2}
      & \multirow{2}{*}{\parbox{6.4cm}{\cpp{SI\_A\_BXsmumu\_highq2} (\cpp{double}):
              \\ Computes the high-$q^2$ forward-backward asymmetry of $B\to X_s\mu^+ \mu^-$.}}
          & \cpp{SuperIso\_modelinfo} & \cpp{A\_BXsmumu\_highq2\_CONV}
                    \\ & & &
                    \\ & & &
  \\ \hline
        \cpp{A\_BXsmumu\_zero}
      & \multirow{2}{*}{\parbox{6.4cm}{\cpp{SI\_A\_BXsmumu\_zero} (\cpp{double}):
              \\ Computes the zero crossing $q^2$ value of the forward-backward asymmetry of $B\to X_s\mu^+ \mu^-$.}}
          & \cpp{SuperIso\_modelinfo} & \cpp{A\_BXsmumu\_zero\_CONV}
          \\ & & &
                    \\ & & &
  \\ \hline
        \cpp{BRBXstautau\_highq2}
      & \multirow{2}{*}{\parbox{6.4cm}{\cpp{SI\_BRBXstautau\_highq2} (\cpp{double}):
              \\ Computes the inclusive high-$q^2$ branching fraction of $B\to X_s\tau^+ \tau^-$.}}
          & \cpp{SuperIso\_modelinfo} &\cpp{BRBXstautau\_highq2\_CONV}
          \\ & & &
          \\ & & &
  \\ \hline
        \cpp{A\_BXstautau\_highq2}
      & \multirow{2}{*}{\parbox{6.4cm}{\cpp{SI\_A\_BXstautau\_highq2} (\cpp{double}):
              \\ Computes the high-$q^2$ forward-backward asymmetry of $B\to X_s\tau^+ \tau^-$.}}
          & \cpp{SuperIso\_modelinfo} & \cpp{A\_BXstautau\_highq2_CONV}
          \\ & & &
          \\ & & &
  \\ \hline
\end{tabular}
}\caption{Observable capabilities of \flavbit that involve Wilson coefficients in their calculation (from \superiso unless otherwise specified).\label{tab:flavourobsWC}}
\end{table*}

\begin{table*}[tp]
\centering
\small{
\begin{tabular}{l|p{6.5cm}|l|l}
  \textbf{Capability}
      & \multirow{2}{*}{\parbox{6.5cm}{\textbf{Function (Return Type)}: \\ \textbf{Brief Description}}} & \textbf{Dependencies}
      & \multirow{2}{*}{\parbox{2cm}{\textbf{Backend\\ requirements}}}
      \\ & & &
  \\ \hline
        \cpp{BKstarmumu_}\metavar{l}\cpp{_}\metavar{m}
      & \multirow{4}{*}{\parbox{6.5cm}{\cpp{SI\_BKstarmumu_}\metavar{l}\cpp{_}\metavar{m} (\cpp{Flav\_KstarMuMu\_obs}):
              \\ Computes all observables associated with $B^0\to K^{*0} \mu^+\mu^-$ in a $q^2$ bin specified by \metavar{l} and \metavar{m}. See caption for details.}}
          & \cpp{SuperIso\_modelinfo} & \cpp{SI\_BKstarmumu\_CONV}
          \\ & & &
          \\ & & &
          \\ & & &
  \\ \hline
      \cpp{AI\_BKstarmumu}
      & \multirow{2}{*}{\parbox{6.5cm}{\cpp{SI\_AI\_BKstarmumu} (\cpp{double}):
              \\ Computes the low-$q^2$ isospin asymmetry of $B\to K^* \mu^+\mu^-$ (in GeV$^2$).}}
          & \cpp{SuperIso\_modelinfo} & \cpp{AI\_BKstarmumu\_CONV}
          \\ & & &
          \\ & & &
  \\ \hline
      \cpp{AI\_BKstarmumu\_zero}
      & \multirow{2}{*}{\parbox{6.5cm}{\cpp{SI\_AI\_BKstarmumu\_zero} (\cpp{double}):
              \\ Computes the zero-crossing $q^2$ value of the isospin asymmetry of $B\to K^* \mu^+\mu^-$.}}
          & \cpp{SuperIso\_modelinfo} & \cpp{AI\_BKstarmumu\_}
          \\ & & & \cpp{\ zero\_CONV}
          \\ & & &
\\ \hline

\end{tabular}
}
\caption{Observable capabilities of \flavbit related to the decay $B^0\to K^{*0} \mu^+\mu^-$.  The indices \metavar{l} and \metavar{m} refer to the edges of the energy bin used in the particular function. The functions and capabilities are named such that \metavar{l},\metavar{m} = 11,25 indicates an energy range of $1.1$--$2.5$ GeV$^2$, and so on. Possible pairs of \metavar{l} and \metavar{m} are (11,25), (25,40), (40,60), (60,80), (15,17) and (17,19); the last two refer to momentum transfer ranges of $15$--$17$--$19$ GeV$^2$.  \label{tab:flavourobsBKstar}}
\end{table*}

\begin{table}[tp]
\centering
\small{
\begin{tabular}{l c}
  \textbf{Name (type)}
      & \textbf{Description}
     \\ \hline
   \cpp{BR} (\cpp{double}) &  branching fraction\\ \hline
   \cpp{AFB} (\cpp{double}) &  forward-backward asymmetry\\ \hline
   \cpp{FL} (\cpp{double}) &  longitudinal fraction\\ \hline
   \cpp{S3} (\cpp{double}) &  $S_3$\\ \hline
   \cpp{S4} (\cpp{double}) &  $S_4$\\ \hline
   \cpp{S5} (\cpp{double}) &  $S_5$\\ \hline
   \cpp{S7} (\cpp{double}) &  $S_7$\\ \hline
   \cpp{S8} (\cpp{double}) &  $S_8$\\ \hline
   \cpp{S9} (\cpp{double}) &  $S_9$\\ \hline
   \cpp{q2\_min} (\cpp{double}) &  $q^2$ bin lower edge \\ \hline
   \cpp{q2\_max} (\cpp{double}) &  $q^2$ bin upper edge \\ \hline
  \end{tabular}
  }
\caption{Observables contained in the \cpp{Flav\_KstarMuMu\_obs} structure.\label{tab:FlavKstarMuMuobs}}
\end{table}

\subsection{Electroweak penguin transitions}
\label{sec:obs_penguin}

Rare semi-leptonic decays of $B$ mesons proceed via flavour-changing neutral currents (FCNCs) in electroweak penguin diagrams, and set stringent constraints on possible contributions from new physics.  \flavbit includes predictions of various FCNC $b\to s$ transitions. These decays are all proportional to the elements $V_{tb}$ and $V_{ts}$ of the CKM matrix.

Rare decays of the type $B\to M \ell^+ \ell^-$, with one meson $M$ in the final state, are sensitive to the Wilson coefficients $C_{9, 10}^{(\prime)}$. In addition, when $M$ is a vector, such as the $K^{*}(892)$, these decays are also sensitive to the Wilson coefficients $C_{7}^{(\prime)}$.

The four-quark operators ($\mathcal{O}_{1\cdots6}$) in the effective Hamiltonian also contribute to the penguin diagrams, resulting in expressions with the same structure as $\mathcal{O}_7$ and $\mathcal{O}_9$. They can therefore be reabsorbed and used to define effective Wilson coefficients $C_7^{\rm eff}$ and $C_9^{\rm eff}$ \cite{Beneke:2001at},
\begin{eqnarray}
C_7^{\rm eff} &=& C_7 -\frac{1}{3} C_3 -\frac{4}{9} C_4 -\frac{20}{3} C_5 -\frac{80}{9} C_6 \;,\\
C_9^{\rm eff} &=&  C_9 + Y(q^2)\;,
\label{effWilson}
\end{eqnarray}
where $Y$ contains the short distance contributions from the four-quark operators \cite{Greub:1994pi,Kruger:1996cv}.

The most accessible of the $B\to M \ell^+ \ell^-$ decays at LHCb are those including final-state muons. The differential decay rate for $B\to M \mu^+\mu^-$, where $M$ is a pseudoscalar, is given at leading order by \cite{Hiller:2003js}:
\begin{align}
\frac{d\Gamma}{dq^2} = \frac{G_F^2 \alpha^2 |V_{tb}V^*_{ts}|^2 \, m_B^3}{(2\pi)^{10}} u(q^2) \Bigg\{ v(q^2) |C_{10} f_+(q^2)|^2 \nonumber\\
+ 4 \frac{m_\mu^2(m_B^2-m_M^2)^2}{q^2 m_B^4}|C_{10} f_0(q^2)|^2\nonumber\\
+ \left|C_9^\text{eff}f_+(q^2) +2 \frac{m_b+m_s}{m_B+m_M} C_7^\text{eff} f_T(q^2)\right|^2\Bigg\}\;,
\end{align}
where $u(q^2)$ and $v(q^2)$ are kinematic factors, and $f_0$, $f_+$ and $f_T$ are $q^2$-dependent form factors.

If $M$ is a vector particle, the $B\to M \ell^+ \ell^-$ decays are completely described by the dilepton invariant mass squared $q^2$ and three angles ($\theta_l, \theta_K$ and $\phi$; see Ref.~\cite{Aaij:2013iag,Gratrex:2015hna} for definitions). Measurements of angular observables of the decays $B^0\to K^*(892)\mu^+\mu^-$ and $B^0_s\to \phi \mu^+\mu^-$ provide a better sensitivity to new physics than measurements of branching fractions. As a function of $q^2$ and the three angles, the differential decay rate for $B^0\to K^{*0}\mu^+\mu^-$ is
\begin{align}
\frac{1}{\Gamma} \frac{\mathrm{d}^{3}(\Gamma +
  \bar{\Gamma})}{\mathrm{d}\cos\theta_{\ell}\,\mathrm{d}\cos\theta_{K}\,\mathrm{d}\phi}
= \frac{9}{32\pi} \left[ \frac{3}{4}(1-{\color{blue}F_\text{L}}) \sin^2 \theta_K \right. \nonumber\\
+\left. {\color{blue}F_\text{L}} \cos^2 \theta_K+ \frac{1}{4}(1-{\color{blue}F_\text{L}}) \sin^2 \theta_K \cos 2\theta_{\ell} \right. \nonumber\\
-\left. {\color{blue}F_\text{L}} \cos^2 \theta_K \cos 2\theta_{\ell} +  {{\color{blue}S_{3}}} \sin^2 \theta_K \sin^{2} \theta_{\ell} \cos 2\phi \right. \nonumber\\
+\left. {\color{blue}S_{4}} \sin 2\theta_K \sin 2\theta_{\ell} \cos\phi + {{\color{blue}S_{5}}} \sin 2\theta_K \sin\theta_{\ell}\cos\phi \right.\nonumber\\
+\left. \frac{4}{3}{\color{blue}A_\text{FB}} \sin^2 \theta_K \cos\theta_{\ell} + {{\color{blue}S_{7}}} \sin 2\theta_K \sin\theta_{\ell} \sin\phi \right. \nonumber\\
+\left.  {\color{blue}S_{8}} \sin 2\theta_K \sin 2\theta_{\ell}\sin\phi + {\color{blue}{S_{9}}} \sin^2 \theta_K \sin^{2}\theta_{\ell} \sin 2\phi ~\right]\;,
\end{align}
where $\bar{\Gamma}$ is the decay rate of the CP conjugate mode. The angular observable $F_\text{L}$ is the longitudinal polarisation fraction of the $K^*$. The other observables are $S_i$, and the forward-backward asymmetry $A_\text{FB}$. The most sensitive experimental analyses assume that there are no scalar contributions (which are constrained by the branching fraction of $B^0_s\to \mu^+\mu^-$), and no tensor contributions.\footnote{Although Ref.\ \cite{Aaij:2015oid} includes measurements free from these assumptions, using the Method of Moments~\cite{Beaujean:2015xea}, the resulting precision is about $15\%$ less than in the likelihood fit.}. This assumption makes it possible to eliminate the observables $S_1^c$, $S_1^s$, $S_2^c$ and $S_2^s$ in favour of a single observable $F_\text{L}$.  The physical observables are sesquilinear combinations of the transversity amplitudes~\cite{Altmannshofer:2008dz},
\begin{eqnarray}
{\color{blue} F_\text{L}} & = &  1-F_T = \frac{A_0^2}{A_{\parallel}^2 + A_{\perp}^2 + A_0^2} \;, \label{fl}\\
{\color{blue} S_3} & = & \frac{1}{2}\frac{A_{\perp}^{L 2}-A_{\parallel}^{L
    2}}{A_{\parallel}^2 + A_{\perp}^2 + A_0^2} + L \to R\;, \\
{\color{blue} S_4} &= & \frac{1}{\sqrt{2}}\frac{\Real (A_0^{L *}
  A_{\parallel}^L )}{A_{\parallel}^2 + A_{\perp}^2 + A_0^2 } + L \to R\;, \\
{\color{blue} S_5} & =& \sqrt{2}\frac{\Real (A_0^{L *}
  A_{\perp}^L )}{A_{\parallel}^2 + A_{\perp}^2 + A_0^2} - L \to R\;, \\
{\color{blue} A_\text{FB}} &=& \frac{8}{3}\frac{\Real
  (A_{\perp}^{L *}A_{\parallel}^{L})}{A_{\parallel}^2 + A_{\perp}^2 + A_0^2}
- L\to R\;, \\
{\color{blue} S_7} &=& \sqrt{2}\frac{\Imag (A_0^{L *}
  A_{\parallel}^L )}{A_{\parallel}^2 + A_{\perp}^2 + A_0^2} + L
\to R\;, \\
{\color{blue} S_8} &=& \frac{1}{\sqrt{2}}\frac{\Imag (A_0^{L *}
  A_{\perp}^L )}{A_{\parallel}^2 + A_{\perp}^2 + A_0^2} + L
\to R\;, \\
{\color{blue} S_9} &=& \frac{\Imag (A_{\perp}^{L *} A_{\parallel}^{L}) }{A_{\parallel}^2 + A_{\perp}^2 + A_0^2 } - L\to R\;.\label{s9}
\end{eqnarray}

The indices $\perp$, $\parallel$ and $0$ refer to the $K^*(892)$ transversity amplitudes, while $L\to R$ refers to the chirality-flipped version of the previous term in each expression.

The amplitudes $A_{\perp,\parallel,0}$ depend on form factors and Wilson coefficients, and can be written at leading order in QCD in the form:
\begin{eqnarray}\label{eq:Amplitudes2}
A_{\perp}^{L,R}&\propto& \Big\{ (C_9^\text{eff}+C_9^{\text{eff}\prime})\mp (C_{10}+C_{10}^{\prime})\frac{V(q^2)}{m_B+m_{K^*}}\nonumber\\
&+&\frac{2m_b}{q^2}(C_7^\text{eff}+C_7^{\text{eff}\prime})T_1(q^2)\Big\}\;,\\
A_{\parallel}^{L,R}&\propto& \Big\{ (C_9^\text{eff}-C_9^{\text{eff}\prime})\mp (C_{10}-C_{10}^{\prime})\frac{A_1(q^2)}{m_B+m_{K^*}}\nonumber\\
&+&\frac{2m_b}{q^2}(C_7^\text{eff}-C_7^{\text{eff}\prime})T_2(q^2)\Big\}\;,\\
A_{0}^{L,R}&\propto& \Big\{\big[ (C_9^\text{eff}-C_9^{\text{eff}\prime})\mp (C_{10}-C_{10}^{\prime})\big]\nonumber\\
&\times& \big[ (m_B^2-m_{K^*}^2-q^2)(m_B+m_{K^*} A_1(q^2)\nonumber\\
&-&\lambda\frac{A_2(q^2)}{m_B+m_{K^*}})\big] + 2m_b(C_7^\text{eff}+C_7^{\text{eff}\prime})\nonumber\\
&\times&\big[(m_B^2+3m_{K^*}^2-q^2)T_2(q^2)\nonumber\\
&-&\frac{\lambda}{m_{B}^2-m_{K^*}^2}T_{3}(q^2)\big]\Big\}\;.
\end{eqnarray}
In the limit of large recoil (low $q^2$), the seven form factors $A_{1,2}$, $T_{1,2,3}$ and $V$ can be replaced by only two form factors $\xi_{\perp}$ and $\xi_{\parallel}$. This makes it possible to write a set of six observables that are independent of form factors in this approximation (see Ref.~\cite{Descotes-Genon:2013vna}).  These are denominated\footnote{Note that for historical reasons the observables $P_{4,5,6}^{\prime}$ \mbox{carry a $^\prime$.}} $P_{i}^{(\prime)}$, with $i\in[1,6]$. Some of these observables were independently proposed by other authors with a different name, e.g. $P_1=A_T^{(2)}$~\cite{Kruger:2005ep}, $P_2=2\times A_{T}^{Re}$~\cite{Becirevic:2011bp}.

The observables $P_i$ can be written as ratios of the observables $F_\text{L}$ and $S_i$, therefore if the full form factors $A_{1,2}$, $T_{1,2,3}$, $V$~\cite{Straub:2015ica} and their correlations are used it is equivalent to using the full set of $P_i$ observables. One of the most interesting measurements in these decays is the observable $P_5^{\prime}$, which shows a deviation with respect to the SM prediction of about 4$\sigma$ in the region $4<q^2/\text{GeV}^2<8$~\cite{Aaij:2013qta,Aaij:2015oid,Abdesselam:2016llu}. The most accredited explanation for this deviation is a reduced $C_9^\text{eff}(q^2)$ Wilson coefficient, but it is not yet clear if this is due to hadronic uncertainties~\cite{Jager:2012uw,Lyon:2014hpa,Ciuchini:2015qxb,Chobanova:2017ghn} or a genuine contribution from new physics~\cite{Descotes-Genon:2013wba,Altmannshofer:2013foa,Hurth:2013ssa,Jager:2014rwa}.
In \flavbit, we incorporate a 10\% theoretical uncertainty (at the amplitude level) into our correlation matrix for $B^0\to K^{*0}\mu^+\mu^-$ observables, to account for errors arising from non-factorisable power corrections~\cite{Hurth:2016fbr}.

As set out in Tables \ref{tab:flavourobsBKstar} and \ref{tab:FlavKstarMuMuobs}, \flavbit can calculate the full suite of observables for $B^0\to K^{*0}\mu^+\mu^-$, in six different $q^2$ bins over the range $1.1 \le q^2/\text{GeV}^2 \le 19.0$.  These are provided by the capabilities \cpp{BKstarmumu_}\metavar{l}\cpp{_}\metavar{m}, where the lower $q^2$ bin edge is denoted by \metavar{l} and the upper edge by \metavar{m}.  The functions with these capabilities return a \cpp{Flav_KstarMuMu_obs} object (Table \ref{tab:FlavKstarMuMuobs}), which contains the overall branching fraction, forward-backward asymmetry and detailed angular observables $F_\text{L}, S_3, S_4, S_5, S_7, S_8$ and $S_9$.  These observables can either be extracted manually from the \cpp{Flav_KstarMuMu_obs} object itself, or output in full via the \GB printer system \cite{gambit} for later analysis.

The angular analysis of $B^0\to K^{*0} \ell^+\ell^-$~\cite{Aaij:2015dea} at much lower momentum transfer ($q^2\lesssim1$ GeV$^2$) can also provide strong constraints, specifically on the coefficients $C_{7}^{(\prime)}$. However, experimental analyses of $B^0\to K^{*0} \mu^+\mu^-$ in this regime are impacted by the assumption that the muon is massless.  We therefore do not include this lower angular bin in \flavbit.

Asymmetries between $B^0$ and $\bar{B^0}$ in ${B^0\to K^{*0}\mu^+\mu^-}$ have also been measured by the LHCb collaboration~\cite{Aaij:2015oid}. These are important for constraining the imaginary parts of a number of Wilson Coefficients.

Another observable useful for isolating the contribution of new physics, owing to its insensitivity to hadronic parameters such as form factors, is the $CP$-averaged $B\to K^*\mu^+\mu^-$ isospin asymmetry \cite{Feldmann:2002iw},
\begin{equation}
\label{ai}
\frac{dA_I}{dq^2} \equiv \frac{d\Gamma_{B^0\to K^{*0}\mu^+\mu^-}/dq^2 - d\Gamma_{B^\pm\to K^{*\pm}\mu^+\mu^-}/dq^2}{d\Gamma_{B^0\to K^{*0}\mu^+\mu^-}/dq^2 + d\Gamma_{B^\pm\to K^{*\pm}\mu^+\mu^-}/dq^2}.
\end{equation}
\flavbit provides the integrated low-$q^2$ asymmetry, corresponding to the integral of Eq.~\ref{ai} over the range $1 \le q^2/\text{GeV}^2 \le 6$ (\cpp{AI_BKstarmumu} in Table \ref{tab:flavourobsBKstar}).  It also computes the zero-crossing of the asymmetry, corresponding to the $q^2$ value where the differential decay rates of $B^0\to K^{*0}\mu^+\mu^-$ and $B^\pm\to K^{*\pm}\mu^+\mu^-$ are equal (\cpp{AI_BKstarmumu_zero} in Table \ref{tab:flavourobsBKstar}).

The measurement of the inclusive branching fraction of $B\to X_s \ell^+\ell^-$ is challenging from the experimental point of view, however has several theory advantages. The differential decay rate at leading order in QCD can be written as~(see Ref. \cite{Huber:2015sra} and references therein):
\begin{eqnarray}
&\displaystyle\frac{d{\cal B}(B\rightarrow X_s\ell^+\ell^-)}{d\hat{s}}=\mathcal{B}\left(B\rightarrow X_cl\bar{\nu}\right)\frac{\alpha^2}{4\pi^2f(z)}\frac{\left|V_{tb}V_{ts}^*\right|^2}{\left|V_{cb}\right|^2}\nonumber\\
&\displaystyle\times(1-\hat{s})^2\sqrt{1-\frac{4m^2_{\ell}}{q^2}}\left\{\left(\left|C_9^\text{eff}\right|^2+\left|C_{10}\right|^2\right)\left(1+2\hat{s}\right)\right.\nonumber\\
&\displaystyle\left. + 4\left|C_7^\text{eff}\right|^2\left(1+\frac{2}{\hat{s}}\right)+12\,\Real\left(C_7^\text{eff}C_9^\text{eff}\right)\right\}\label{BR_BXsll}
\end{eqnarray}
where $\hat{s}\equiv q^2/m^2_b$, $z=m_c^2/m_b^2$ and
\begin{equation}
 f(z)=1-8z+8z^3-z^4-12z^2\ln z\;.
\label{fz}
\end{equation}
The inclusive and differential branching fractions of $B\to X_s \ell^+\ell^-$ were measured at $B$ factories~\cite{Aubert:2004it,Iwasaki:2005sy,Sato:2014pjr,Lees:2013nxa}.

As detailed in Table \ref{tab:flavourobsWC}, \flavbit computes predictions for ${\cal B}(B\to X_s \mu^+\mu^-)$, integrated over both high and low $q^2$ ranges (capabilities \cpp{BRBXsmumu_highq2} and \cpp{BRBXsmumu_lowq2}).  It also computes the branching fraction at high $q^2$ for the equivalent process with $\tau$ leptons in the final state, ${\cal B}(B\to X_s \tau^+\tau^-)$ (capability \cpp{BRBXstautau_highq2}).

A complementary $B\to X_s \ell^+\ell^-$ angular observable is the forward-backward asymmetry $A_{\text{FB},B\to X_s \ell^+\ell^-}$, defined differentially with respect to $\hat{s}$ as
\begin{equation}
A_{\text{FB},B\to X_s \ell^+\ell^-}(\hat{s}) \equiv \int_0^1 \frac{d{\cal B}(\hat{s},z)}{d\hat{s}dz} - \int_{-1}^0 \frac{d{\cal B}(\hat{s},z)}{d\hat{s}dz},
\end{equation}
where $z$ is the cosine of the forward angle.  \flavbit computes the $B\to X_s \mu^+\mu^-$ integrated forward-backward asymmetry at both low and high $q^2$ (capabilities \cpp{A_BXsmumu_highq2} and \cpp{A_BXsmumu_lowq2}), along with the zero-crossing of the asymmetry, corresponding the $q^2$ value for which the asymmetry vanishes (\cpp{A_BXsmumu_zero}).  It also predicts the asymmetry of the equivalent process involving $\tau$ leptons at high $q^2$ (capability \cpp{A_BXstautau_highq2}).

The decay $B_s\to \phi \mu^+\mu^-$ is described by the same formalism as $B\to K^*\mu^+\mu^-$. However, while the latter is a self-tagging decay, i.e. the flavour of the $B$ meson at decay time can be inferred by the charge of the kaon coming from the decay of the $K^*(892)$, this is not the case for the $B_s\to \phi \mu^+\mu^-$. This implies that when averaging between $B_s$ and $\bar{B}_s$, some terms of the angular distributions (including $P_5^{\prime}$) vanish.  The branching ratios of both $B_s\to \phi \mu^+\mu^-$ and the related decay $B^+\to K^+\mu^+\mu^-$ are sensitive to BSM physics, mainly via the Wilson coefficients $C_{9}^{(\prime)}$ and $C^{(\prime)}_{10}$.  The measurement of the branching fraction of $B_s\to \phi \mu^+\mu^-$ by the LHCb experiment~\cite{Aaij:2015esa} is also in tension with respect to SM predictions.  We do not include these channels directly in \flavbit, because to do so rigorously would require the ability to recompute model-dependent BSM contributions to theoretical uncertainties.  This is a capability that we anticipate including in a future version of \flavbit.

In addition, angular measurements of the decay $B^0\to K\pi \mu^+\mu^-$ outside the $K^*(892)$ resonance have been recently performed~\cite{Aaij:2016kqt}, however we do not yet have enough knowledge of the different $K^*$ resonances in that region of $K\pi$ invariant mass to interpret the result in terms of Wilson coefficients~\cite{Das:2014sra}. For this reason, the decays $B^0\to K\pi \mu^+\mu^-$ outside the $K^{*0}(892)$ are not yet implemented in \flavbit.

Lepton flavour universality in $b\to s$ transitions has also been tested by measuring the ratio $R_K=\frac{{\cal B}(B^+\to K^+\mu^+\mu^-)}{{\cal B}(B^+\to K^+e^+e^-)}$. A tension corresponding to $2.6\sigma$ was observed~\cite{Aaij:2014ora}. Contrary to the anomalies in the aforementioned $b \to s\ell\ell$ transitions, the tension in $R_K$ cannot be explained by hadronic uncertainties. Accommodating lepton flavour non-universality within the effective Hamiltonian framework of Eq.~\ref{physical_basis} requires splitting operators $\mathcal{O}_9^{(')}$ and $\mathcal{O}_{10}^{(')}$ into separate effective operators for different leptons. In the context of this expanded treatment, the so-called flavour anomalies in rare decays seem to form a coherent pattern, with a reduction of about 25\% observed in the muonic $C_9$ Wilson coefficient relative to the SM prediction. In general these scenarios are not easy to accommodate within the MSSM, although a global agreement at the 2$\sigma$ level is still possible~\cite{Mahmoudi:2014mja}.  Presently, \flavbit does not deal with violations of lepton flavour universality, so $R_K$ is not yet included as an observable.

\subsection{Rare purely leptonic decays}
\label{sec:obs_rare}

Like its penguin counterparts $B\to X \ell^+ \ell^-$, the rare leptonic decay $B^0_{s}\to \ell^+ \ell^-$ also probes the FCNC $b\to s$ transition, and is proportional to the CKM entries $V_{tb}$ and $V_{ts}$.  Similarly, $B^0\to \ell^+ \ell^-$ probes $b\to d$ and is proportional to $V_{tb}$ and $V_{td}$.  These are rather clean channels from the theoretical perspective, as the main uncertainty comes only from the meson decay constant, which can be calculated in lattice QCD. The branching fraction of these decays is
\begin{align}
\label{eq:BsmmBR}
{\cal B}(B_q^0\to\ell^+\ell^-) = \frac{G_F^2\alpha^2}{64\pi^3}f_{B_q}^2\tau_{B_q}m_{B_q}^3\left|V_{tb}V_{tq}^*\right|^2\nonumber\\
\times\sqrt{1-\frac{4m_{\ell}^2}{m_{B_q}^2}} \Bigg\{ \left( 1 -\frac{4m_{\ell}^2}{m_{B_q}^2}\right)  \left|C_{Q1}-C_{Q1}^{\prime}\right|^2\nonumber\\
+\left|(C_{Q2}-C_{Q2}^{\prime})+2\frac{m_{\ell}}{m_{B_q}}(C_{10}-C_{10}^{\prime})\right|^2 \Bigg\}\;.
\end{align}

Because the $B$ meson is a pseudoscalar, these decays are helicity-suppressed, in addition to the GIM suppression. Therefore, in the SM and in all lepton-flavour-universal $V\pm A$ models, the ratio of the branching fractions for different leptons is given by:
\begin{equation}
\frac{{\cal B}(B_q\to \ell_1^+ \ell_1^- )}{{\cal B}{(B_q\to \ell_2^+ \ell_2^- )}}=\frac{m_{1}^2}{m_2^2}\;,
\end{equation}
where $m_{1(2)}$ is the mass of the lepton $\ell_{1(2)}$.  These decays set strong constraints on models with extended Higgs sectors such as the 2HDM, as scalar contributions would alleviate the helicity suppression. Such decays are also sensitive to new bosons with $V\pm A$ couplings (e.g. $W^{\prime}$ and $W_{R}$), which would modify the Wilson coefficients $C_{10}^{(\prime)}$ of the SM.

\flavbit has the capability to compute the branching fraction for $B^0\to\mu^+\mu^-$ (\cpp{Bmumu} in Table \ref{tab:flavourobsWC}), as well as for ($CP$-averaged) $B_s$ decays to $e^+e^-$ and $\mu^+\mu^-$ (\cpp{Bsee_untag} and \cpp{Bsmumu_untag}).  The latter can also be obtained in the MSSM and SM from \FH via the \cpp{FH_FlavourObs} capability (see Tables \ref{tab:flavourobswoWC} and \ref{tab:fhFlavourObs}).

\begin{table}[tp]
\centering
\small{
\begin{tabular}{l|l}
  \textbf{Name}
      & \textbf{Description}
     \\ \hline
   \cpp{Bsg\_MSSM} (\cpp{fh\_real})    & Total inclusive branching fraction of \\
                                       & $B\to X_s\gamma$ in the MSSM \\\hline
   \cpp{Bsg\_SM} (\cpp{fh\_real})      & Total inclusive branching fraction of \\
                                       & $B\to X_s\gamma$ in the SM \\\hline
   \cpp{DeltaMs\_MSSM} (\cpp{fh\_real})& $B^0_s-\bar{B^0_s}$ mass difference \\
                                       & in the MSSM\\ \hline
   \cpp{DeltaMs\_SM} (\cpp{fh\_real})  & $B^0_s-\bar{B^0_s}$ mass difference \\
                                       & in the SM\\ \hline
   \cpp{Bsmumu\_MSSM} (\cpp{fh\_real}) & Branching fraction of \\
                                       & $B^0_s\to \mu^+ \mu^-$ in the MSSM\\ \hline
   \cpp{Bsmumu\_SM} (\cpp{fh\_real})   & Branching fraction of \\
                                       & $B^0_s\to \mu^+ \mu^-$ in the SM\\ \hline
  \end{tabular}
  }
\caption{Flavour observables contained in the \cpp{fh\_FlavourObs} structure obtained from \FH.\protect\footnotemark \label{tab:fhFlavourObs}}
\end{table}

\subsection{Other flavour observables}
\label{sec:obs_other}

Other observables included in \flavbit are $B\to X_s\gamma$, the ratio $R_{\mu}=\frac{{\cal B}(K\to \mu \nu_\mu)}{{\cal B}(\pi \to \mu \nu_\mu)}$, and the meson mixing $\Delta M_s$.

Radiative decays of $B$ mesons are important to constrain the electromagnetic operator and the corresponding Wilson coefficients $C_7^{(\prime)}$. The main constraint comes from the measurement of the inclusive decay $B\to X_s \gamma$~\cite{Bertolini:1990if,Misiak:2006zs}. The prediction of this branching fraction is relatively clean, and benefits from the Heavy Quark Expansion in the same way as the $B\to X_s \ell^+ \ell^-$ process.

The branching ratio can be written at leading order as
\begin{equation}
{\cal B}(\bar{B} \to X_s \gamma)= {\cal B}(\bar{B} \to X_c e \bar{\nu})_{\rm exp} \left| \frac{ V^*_{ts} V_{tb}}{V_{cb}} \right|^2 \frac{6 \alpha}{\pi C} \left|C_7^\text{eff}\right|^2,
\end{equation}
where ${\cal B}(\bar{B} \to X_c e \bar{\nu})_{\rm exp}$ is the experimentally-measured value of the branching fraction for $\bar{B} \to X_c e \bar{\nu}$, and
\begin{equation}
C = \left| \frac{V_{ub}}{V_{cb}} \right|^2 \frac{{\cal B}(\bar{B} \to X_c e \bar{\nu})}{{\cal B}(\bar{B} \to X_u e \bar{\nu})}\;.
\end{equation}
This measurement sets constraints on the charged Higgs mass and couplings of the 2HDM~\cite{Mahmoudi:2009zx,Hermann:2012fc,Misiak:2015xwa,Misiak:2017bgg}. In addition, these measurements constrain models with additional neutral gauge bosons such as the $Z^{\prime}$~\cite{Greljo:2015mma}. \flavbit implements this observable as \cpp{bsgamma} (Table \ref{tab:flavourobsWC}), and within the \cpp{FH_FlavourObs} capability (see Tables \ref{tab:flavourobswoWC} and \ref{tab:fhFlavourObs}).

\footnotetext{Note that the branching fraction of $B\to X_s\gamma$ is ill-defined for $E_\gamma\to0$, due to the IR divergence associated with soft photon emission.  Although the adopted cutoff on $E_\gamma$ is unspecified in \FH, $\mathcal{B}(B\to X_s\gamma)$ here appears to follow the definition of `total' advocated in Ref.\ \cite{Kagan:1998ym}, with  $E_\gamma > m_b/10 \sim 0.4$\,GeV.}

The exclusive decays $B\to K^* \gamma$ and $B_s\to \phi \gamma$ also constrain the coefficients $C_{7}^{(\prime)}$, but their impact is not yet competitive with the inclusive one. However, the inclusive decays can only constrain the sum of $|C_7|^2$ and $|C_7^{\prime}|^2$. The best constraint on the right-handed current $C_7^{\prime}$ contribution presently comes from the angular analysis of $B^0\to K^{*0} \ell^+\ell^-$ at low $q^2$ (see Sec.\ \ref{sec:obs_penguin}). In \flavbit, we provide the $CP$-averaged isospin asymmetry of $B\to K^* \gamma$ decays~\cite{Kagan:2001zk},
\begin{equation}
\Delta_0 \equiv \frac{\Gamma(\bar{B}^0\to \bar{K}^{*0}\gamma) - \Gamma(B^\pm\to K^{*\pm}\gamma)}{\Gamma(\bar{B}^0\to \bar{K}^{*0}\gamma) + \Gamma(B^\pm\to K^{*\pm}\gamma)},
\end{equation}
as a calculable observable, as it can receive contributions from charged Higgs bosons and any other new fields with similar quantum numbers (such as charginos in supersymmetry)~\cite{Ahmady:2006yr}. The predicted asymmetry can be accessed via capability \cpp{delta0} (Table \ref{tab:flavourobsWC}).

The leptonic decays of $K$ and $\pi$ mesons are also sensitive to the existence of charged Higgs bosons~\cite{Gonzalez-Alonso:2016etj}.
\flavbit computes the ratio \cite{Antonelli:2008jg}
\begin{eqnarray}
R_{\mu} &=& \dfrac{{\cal B}(K \to \mu \nu_\mu)}{{\cal B}(\pi \to \mu \nu_\mu)}\nonumber\\
&=& \left(1+\delta_{\rm em}\right)\frac{\tau_K}{\tau_\pi}\left|\frac{V_{us}}{V_{ud}} \right|^2 \frac{f^2_K}{f^2_\pi} \frac{m_K}{m_\pi}\left(\frac{1-m^2_\ell/m_K^2}{1-m^2_\ell/m_\pi^2}\right)^2  \nonumber\\
&&\times \left[1-\frac{m^2_{K^+}}{M^2_{H^+}}\left(1 - \frac{m_d}{m_s}\right)\frac{\tan^2\beta}{1+\epsilon_0\tan\beta}\right]^2,
\end{eqnarray}
which has a smaller theoretical uncertainty than the individual decays.  Here $\delta_{\rm em} = 0.0070 \pm 0.0035$ is a long-distance electromagnetic correction factor.  We also consider the quantity $R_{\mu 23}$ \cite{Antonelli:2008jg},
\begin{eqnarray}
R_{\mu 23} &=&\left| \frac{V_{us}(K_{\ell 2})}{V_{us}(K_{\ell 3})} \times \frac{V_{ud}(0^+ \to 0^+)}{V_{ud}(\pi_{\ell 2})} \right|\nonumber\\
&=&\left|1-\frac{m^2_{K^+}}{M^2_{H^+}}\left(1 - \frac{m_d}{m_s}\right)\frac{\tan^2\beta}{1+\epsilon_0\tan\beta}\right|\;,
\label{eq:rmu23}
\end{eqnarray}
where $\ell_i$ refers to leptonic decays with $i$ particles in the final state, and $0^+ \to 0^+$ corresponds to nuclear beta decay. These are provided by capabilities \cpp{Rmu} and \cpp{Rmu23}, respectively, and the relevant functions are detailed in Table \ref{tab:flavourobswoWC}.

It is well known that neutral meson systems are characterised by a rich phenomenology. In general, eigenstates of flavour are not eigenstates of mass, causing neutral mesons to oscillate.  The parameters governing oscillations are the difference in mass between the heavy and light eigenstates $\Delta M = M_{H}-M_{L}$ and the difference in their decay widths $\Delta \Gamma = \Gamma_H-\Gamma_L$.
While in the neutral kaon system the difference in lifetime is very large, so we denote the two states `short' ($K^0_S$) and `long' ($K^0_L$), in the neutral $B$ system $\Delta \Gamma \ll \Delta M$, so it is more suitable to call them `heavy' and `light'. The oscillation frequency is related to the difference in mass $\Delta M_q$, which for the neutral $B$ meson is
\begin{equation}
\Delta M_q =\frac{G_F^2}{6\pi^2}\eta_B m_{B_q}(\hat{B}_qf^2_{B_q})M_W^2S_0(x_t)|V_{tq}|^2\;,
\end{equation}
where $\hat{B}_q$ is the renormalisation-group-invariant parameter, $f_{B_q}$ is the $B_q$ decay constant and $S_0(x_t)$ is a simple function of the top mass.
The hadronic parameter $f_{B_q}$ is the same factor that appears in the branching fraction of $B_q\to \ell^+ \ell^-$ decays (Eq.~\ref{eq:BsmmBR}). The branching fractions and mass differences are therefore related as \cite{Buras:2002vd}
\begin{equation}
\frac{{\cal B}(B_s^0\to\ell^+\ell^-)}{{\cal B}(B^0\to\ell^+\ell^-)}=\frac{\hat{B}_s}{\hat{B}_d}\frac{\tau(B^0_s)}{\tau(B^0)}\frac{\Delta M_s}{\Delta M_d}\;.
\end{equation}
In \flavbit, $\Delta M_s$ can be obtained in either the SM or MSSM, via the \cpp{FH_FlavourObs} capability (see Tables \ref{tab:flavourobswoWC} and \ref{tab:fhFlavourObs}).

\begin{table*}[tp]
\centering
\small{
\begin{tabular}{lp{12cm}}
  \textbf{Name} & \textbf{Description}
   \\ \hline
   \yaml{name} & Unique name of a given measurement\\ \hline
   \yaml{islimit} & Flag that indicates if the measurement is in the form of an upper limit (\yaml{true}) or a measurement (\yaml{false}) \\ \hline
   \yaml{exp_value} & The experimental measurement (if \yaml{islimit = false}) or limit (if \yaml{islimit = true}) \\ \hline
   \yaml{exp_stat_error} & $1\sigma$ uncorrelated statistical uncertainty on the experimental measurement or limit\\ \hline
   \yaml{exp_sys_error} & $1\sigma$ uncorrelated systematic uncertainty on the experimental measurement or limit\\ \hline
   \yaml{exp_source} & The source of the experimental value and uncertainties\\ \hline
   \yaml{th_error} & $1\sigma$ uncorrelated theoretical uncertainty\\ \hline
   \yaml{th_error_type} & Flag indicating whether the theory error is multiplicative (\yamlvalue{M}) or additive (\yamlvalue{A}).\\ \hline
   \yaml{th_source} & The source of the theoretical uncertainty\\ \hline
   \yaml{correlation} &  Sub-section with correlations of the experimental measurement/limit to other experimental measurements/limits: \\
   & \begin{tabular}{ll}
     \yaml{name} & Name of another measurement with which this one is correlated \\
     \yaml{value} & Correlation matrix entry relating the two measurements \\
     \yaml{name} & Name of a third measurement with which this one is correlated \\
     \yaml{value} & etc
   \end{tabular}
   \\ \hline
 \end{tabular}
 }
\caption{Parameters of a single experimental entry in the \flavbit \YAML database.\label{tab:names}}
\end{table*}

\begin{table*}[tp]
\centering
\small{
\begin{tabular}{p{5cm}p{11.5cm}}
  \textbf{Name} & \textbf{Description} \\ \hline
  \cpp{int read\_yaml(str name)} & Reads an entire \YAML database file \cpp{name} into memory. \\ \hline
  \cpp{void read\_yaml\_measurement(str} \mbox{\ \cpp{name, str measurement\_name)}} & Extracts a single measurement \cpp{measurement\_name} from the \YAML database file \cpp{name}. \\ \hline
  \cpp{void debug\_mode(bool debug)} & Turns on (\cpp{debug = true}) or off (\cpp{debug = false}) printing of all parameters. \\ \hline
  \cpp{void create\_global\_corr()} & Constructs a total correlation matrix from all measurements read in.\\ \hline
  \cpp{void print\_corr\_matrix()} & Prints the constructed correlation matrix. \\ \hline
  \cpp{void print\_cov\_matrix()} & Prints the corresponding covariance matrix.\\ \hline
  \cpp{void print\_cov\_inv\_matrix()} & Prints the inverse of the covariance matrix.\\ \hline
  \cpp{matrix(}\metavar{n}\cpp{,}\metavar{n}\cpp{) get_cov()} & Returns the experimental covariance matrix covering all measurements read in.\\ \hline
  \cpp{matrix(}\metavar{n}\cpp{,1) get_exp_value()} & Returns the central experimental values for all measurements read in. \\ \hline
  \cpp{matrix(}\metavar{n}\cpp{,1) get_th_err()} & Returns the central (uncorrelated) theory error for each of the measurements read in. \\ \hline
\end{tabular}
}
\caption{Important methods of the \flavbit\ \cpp{Flav\_reader} class.  Here \cpp{str} is an alias for \cpp{std::string}, \metavar{n} is the number of measurements so far read in by the \cpp{Flav\_reader} instance, and \cpp{matrix(}\metavar{x},\metavar{y}\cpp{)} is an \metavar{x} $\times$ \metavar{y} \cpp{boost::numeric::ublas::matrix<double>}. \label{tab:flav_reader} \label{tab:flavreader}}
\end{table*}

\begin{table*}[tp]
\centering
\small{
\begin{tabular}{l|p{10.5cm}|l}
  \textbf{Capability}     & \multirow{2}{*}{\parbox{10.5cm}{\textbf{Function (Return Type)}: \\ \textbf{Brief Description}}} & \textbf{Dependencies} \\ & &  \\ \hline
  \cpp{SL\_M}   & \multirow{2}{*}{\parbox{10.5cm}{\cpp{SL\_measurements} (\cpp{predictions\_measurements\_covariances}):
             \\ Tree-level leptonic and semi-leptonic decay predictions, measurements and covariances.}}
          & \cpp{RD} \\ & & \cpp{RDstar}\\ & & \cpp{BDmunu}\\ & & \cpp{BDstarmunu} \\ & & \cpp{Btaunu} \\ & & \cpp{Dstaunu}\\ & & \cpp{Dsmunu}\\ & & \cpp{Dmunu}\\ \hline
  \cpp{SL\_LL}   & \multirow{2}{*}{\parbox{10.5cm}{\cpp{SL\_likelihood} (\cpp{double}):
             \\ Log-likelihood for tree-level leptonic and semi-leptonic decays.}}
          & \cpp{SL\_M} \\ & & \\ \hline
  \cpp{b2sll\_M}   & \multirow{2}{*}{\parbox{10.5cm}{\cpp{b2sll\_measurements} (\cpp{predictions\_measurements\_covariances}):
             \\ Electroweak penguin decay predictions, measurements and covariances.}}
          & \cpp{BKstarmumu\_11\_25} \\  & & \cpp{BKstarmumu\_25\_40}  \\  & &  \cpp{BKstarmumu\_40\_60}  \\  & & \cpp{BKstarmumu\_60\_80}  \\  & & \cpp{BKstarmumu\_15\_17}  \\  & &  \cpp{BKstarmumu\_17\_19}\\ \hline
  \cpp{b2sll\_LL}   & \multirow{2}{*}{\parbox{10.5cm}{\cpp{b2sll\_likelihood} (\cpp{double}):
             \\ Log-likelihood for electroweak penguin decays, including angular observables.}}
          & \cpp{b2sll\_M} \\ & & \\ \hline
  \cpp{b2ll\_M}   & \multirow{2}{*}{\parbox{10.5cm}{\cpp{b2ll\_measurements} (\cpp{predictions\_measurements\_covariances}):
             \\ Rare purely leptonic decay predictions, measurements and covariances.}}
          & \cpp{Bsmumu\_untag} \\  & & \cpp{Bmumu} \\   \hline
  \cpp{b2ll\_LL}   & \multirow{2}{*}{\parbox{10.5cm}{\cpp{b2ll\_likelihood} (\cpp{double}):
             \\ Log-likelihood for rare purely leptonic decays.}}
          & \cpp{b2ll\_M} \\ & & \\  \hline
  \cpp{b2sgamma\_LL}   & \multirow{2}{*}{\parbox{10.5cm}{\cpp{b2sgamma_likelihood} (\cpp{double}):
              \\ Log-likelihood for the branching fraction of $B \to X_s \gamma$.}}
          &  \cpp{bsgamma}  \\ & &  \\  \hline
  \cpp{deltaMB\_LL}   & \multirow{2}{*}{\parbox{10.5cm}{\cpp{deltaMB_likelihood} (\cpp{double}):
              \\ Log-likelihood for $B$ meson mass asymmetries.}}
          &  \cpp{deltaMs}  \\ & &  \\  \hline
\end{tabular}
}
\caption{Likelihood capabilities of \flavbit.  All measurement functions (capabilities ending in \cpp{_M}) return experimental and theoretical central values, as well as experimental and theoretical covariance matrices.}\label{tab:likelihoods}
\end{table*}

\section{Likelihoods}
\label{sec:like}

After calculating the observables described in Section~\ref{sec:obs}, \flavbit can be used to compute likelihoods based on a comparison of the predictions with current experimental measurements.

The experimental results and theoretical errors are stored in a \YAML database.  Taking the branching fraction of $B_s \to \mu^+\mu^-$ as an example, the \flavbit database entry is
\begin{lstyaml}
- name: BR_Bs2mumu
  islimit: false
  exp_value: 3.0e-9
  exp_stat_error: 0.6e-9
  exp_sys_error: 0.25e-9
  exp_source: 1703.05747
  th_error: 0.1
  th_error_type: M
  th_error_source: 1208.0934
  correlation:
  - name: NONE
\end{lstyaml}
The individual fields available in such entries are described in detail in Table~\ref{tab:names}.  Note in particular that the theory error may be given either as a fraction, as in this example, or as an absolute value.   The \cpp{Flav\_reader} object is responsible for reading the experimental results and theoretical errors, and calculating the resulting covariance matrix. Table~\ref{tab:flavreader} describes its specific functions.

We consider correlated theoretical and experimental uncertainties separately, building two covariance matrices and assuming linear correlations for both.  In the case of asymmetric uncertainties, we symmetrise the errors by taking the mean of the upper and lower uncertainties.  \flavbit constructs the experimental covariance matrix directly from the \yaml{exp_stat_error}, \yaml{exp_sys_error} and \yaml{correlation} entries in its \YAML database (Table~\ref{tab:names} and example above).  It takes the \yaml{th_error} entries in the \YAML database and uses them to populate the diagonal of the theory covariance matrix.  It determines the off-diagonal terms on a case-by-case basis in each likelihood function, in order to make it possible for different likelihood functions to adjust the correlations according to whether different nuisance parameters are scanned over directly, or should be included via the correlation matrix.\footnote{Users of \flavbit should be aware of a potential pitfall arising from this arrangement.  The theory uncertainties and correlations that we include in the current release and describe in this paper already incorporate uncertainties on input parameters such as form factors, decay constants, SM masses and couplings, and in particular, CKM matrix entries.  The SM masses and couplings are sufficiently well constrained that any error term dominated by them can be safely neglected, and generally is in \flavbit, seeing as they can be easily varied within \GB as nuisance parameters.  On the other hand, CKM elements are substantial and dominant contributors to the error budget of some processes.  The current likelihoods in \flavbit should therefore \textit{not} be employed in any scan where CKM elements are varied as nuisance parameters, without first carefully considering which likelihood terms already include their impact, and either removing those observables from the fit, or reducing the theory errors accordingly.}

\flavbit builds the full covariance matrix by summing the experimental and theoretical covariance matrices.  If an observable and its measurements are uncorrelated with other observables, the resulting uncertainty then becomes simply the sum in quadrature of the theoretical and experimental errors.

We determine likelihoods for flavour observables under the assumption of correlated Gaussian errors and Wilks' Theorem, taking (twice) the final log-likelihood to be $\chi^2$ distributed. This gives
\begin{equation}
\log \mathcal{L} = -\frac{1}{2} \chi^2 = -\frac{1}{2} \sum_{i,j=1}^N (y_i - x_i) V^{-1}_{ij} (y_j-x_j),
\end{equation}
where $x_i$ is the experimental measurement of the $i$th observable, $y_i$ is the $i$th theory prediction and $V^{-1}$ is the inverse of the full covariance matrix.

\flavbit contains five different likelihood functions. These correspond to different likelihood classes within which observables might be correlated.
\begin{description}
\item \cpp{SL_likelihood}: tree level leptonic and semi-leptonic $B$ and $D$ decays ($B^\pm \to \tau \nu$, $D_{(s)}^\pm \to \ell \nu_\ell$, $B\to D^{(\ast)} \ell \nu_\ell$)
\item \cpp{b2sll_likelihood}: electroweak penguin decays ($B\to X_s \ell^+\ell^-$)
\item \cpp{b2ll_likelihood}: rare purely leptonic $B$ decays ($B^0_{(s)}\to \ell^+\ell^-$)
\item \cpp{b2sgamma_likelihood}: rare radiative $B$ decays ($B\to X_s\gamma$)
\item \cpp{deltaMB_likelihood}: $B$ meson mass asymmetries
\end{description}
The likelihood functions, their capabilities and dependencies are given in Table~\ref{tab:likelihoods}.  In the following subsections, we give details of the experimental data included in each.

\subsection{Tree-level leptonic and semi-leptonic likelihood}

We take the branching fractions of the decays $B\to D^{(*)}\ell \nu_\ell$ from the PDG \cite{Olive:2016xmw}, which combines results from many experiments but is dominated by the contributions from BaBar \cite{Aubert:2007qw,Aubert:2009ac} and Belle \cite{Dungel:2010uk,Glattauer:2015teq}.

BaBar \cite{Lees:2012xj,Lees:2013uzd} and Belle \cite{Huschle:2015rga,Sato:2016svk,Hirose:2016wfn} also recently measured the ratios $R_{D^{(*)}}\equiv{\cal B}(B\to D^{(*)}\tau \nu_\tau)/{\cal B}(B\to D^{(*)}\ell \nu_\ell)$. LHCb also measured $R_{D^*}$ for the muonic final state \cite{Aaij:2015yra}. The average of these measurements, assuming lepton flavour universality between muons and electrons, has been computed by the HFAG collaboration~\cite{Amhis:2016xyh,HFAG17_moriond} and is included in \flavbit:
\begin{eqnarray}
R_D &=&0.403\pm 0.040\pm 0.024\;, \\
R_{D^*} &=&0.310\pm 0.015\pm 0.008\;.
\end{eqnarray}
Compared to the SM predictions of $R_D=0.300\pm 0.008$~\cite{Na:2015kha} and $R_{D^*}=0.252\pm 0.003$~\cite{Fajfer:2012vx}, a total discrepancy of about $4\sigma$ is observed.  We take the experimental correlation between $R_D$ and $R_{D^*}$, arising from common systematics in the measurements, from Ref.\ \cite{Amhis:2016xyh}. The theory uncertainties are considered uncorrelated; we take these from Refs.~\cite{PhysRevD.85.094025,Lattice:2015tia}.

In addition to $R_D$ and $R_{D^*}$, we also explicitly include in the likelihood the decays $B \to D^{(*)} \mu \nu$, adopting the experimental values from the PDG~\cite{Olive:2016xmw}. Taken with $R_D$ and $R_{D^*}$, this set of four likelihood terms constitutes a complete basis for the models of lepton non-universality. The theory errors for the $B \to D^{(*)} \mu \nu$ branching fractions are dominated by form factors \cite{Sakaki:2013bfa,Na:2015kha}. Performing a detailed error analysis with \superiso gives a theoretical uncertainty of 9\% for $B \to D \mu \nu$ and 11\% for $B \to D^* \mu \nu$.

Experiments have not measured any correlation between the muonic and tauonic modes of the decays contributing to $R_{D^{(*)}}$.  However, the theory systematics are strongly correlated; in our analysis with \superiso, we find anti-correlations at the level of 55\% for $B \to D \mu \nu$ and $R_D$, and 62\% for $B \to D^{(*)} \mu \nu$ and $R_{D^*}$.  These data are all included in the \flavbit likelihood.

For $B^\pm\to\ell\nu_\ell$, \flavbit uses experimental measurements from the PDG~\cite{Olive:2016xmw},
\begin{eqnarray}
{\cal B}(B^+ \to \tau^+\nu_{\mu})&=&(1.09\pm 0.24)\times 10^{-4}\;.
\end{eqnarray}
This average is dominated by results from the BaBar~\cite{Lees:2012ju,Aubert:2009wt} and Belle~\cite{Adachi:2012mm,Hara:2010dk} experiments, and is in agreement with the SM. We take this measurement to be uncorrelated with all other measurements. The dominant theoretical uncertainty comes from the CKM element $V_{ub}$.  The present uncertainty on this element is 9.5\% \cite{Olive:2016xmw}, giving an overall theoretical uncertainty of $19\%$.

For the branching fractions of the $D^\pm_{(s)}$ decays $D^\pm \to \mu \nu_\mu$, $D_s^\pm \to \tau \nu_\tau$ and $D_s^\pm \to \mu \nu_\mu$, we adopt the experimental values of the PDG~\cite{Olive:2016xmw}.  (\flavbit does not include $D^\pm \to \tau \nu_\tau$ as an observable, as its decay branching fraction has not yet been measured.) The theory errors on the $D_{(s)}^\pm$ decays are dominated by the knowledge of the decay constant of the corresponding charmed mesons, $f_D$ and $f_{D_{s}}$. This leads to a theoretical uncertainty on the branching fractions of 3\% for $D^\pm$ decays and 2\% for $D_s^\pm$ decays \cite{Olive:2016xmw}.

As shown in Table \ref{tab:likelihoods}, \flavbit collects together into \cpp{SL_M} the measured values, experimental correlations, theoretical predictions and theory uncertainties for $B^\pm\to\ell\nu_\ell$, the four $B \to D^{(*)} \ell \nu_\ell$ observables, and the three $D_{(s)}^\pm$ decays.  This fills the only dependency of the final tree-level leptonic and semi-leptonic likelihood, which can be accessed via capability \cpp{SL_LL}.

\subsection{Electroweak penguin likelihood}
The electroweak penguin likelihood in \flavbit is calculated using the angular observables of the $B^0\to K^{*0}\mu^+\mu^-$ decay, as measured by LHCb~\cite{Aaij:2015oid} in dimuon invariant mass squared bins of (1.1, 2.5), (2.5, 4), (4, 6), (6, 8), (15, 17) and (17, 19)\,GeV$^2$. The bin (11, 12.5)\,GeV$^2$ cannot be used in the likelihood, as the relative phase between the charmonium resonances in this bin and the non-resonant decay is not currently known. We do not implement the measurements of Belle \cite{Abdesselam:2016llu}, as their contribution to the likelihood is negligible compared to the LHCb measurement.  ATLAS and CMS have also very recently presented preliminary Run I measurements of the $B^0\to K^{*0}\mu^+\mu^-$ angular observables \cite{ATLAS-CONF-2017-023, CMS:2017ivg}; these data will be included in a future release of \flavbit.

For each $q^2$ bin, the \flavbit likelihood includes components arising from \cpp{FL}, \cpp{S3}, \cpp{S4}, \cpp{S5}, \cpp{AFB}, \cpp{S7}, \cpp{S8} and \cpp{S9}.  It accounts for experimental correlations between these measurements within each bin, but assumes that measurements are not correlated across $q^2$ bins, as the uncertainty is dominated by the statistical component. The full correlation matrices within each bin are available publicly from LHCb \cite{Aaij:2015oid} and included in the \flavbit \YAML database. We include theory-induced correlated uncertainties between different angular observables for the same $q^2$ range from Ref.~\cite{Hurth:2016fbr,Mahmoudi:2016mgr}.

The branching fractions for $B^0\to K^{*0}\mu^+\mu^-$ decays are not part of the electroweak penguin likelihood in \flavbit \textsf{1.0.0}, but are slated for inclusion in a future version, following the next update from LHCb.  The isospin asymmetry of the $B^0\to K^{*0}\mu^+\mu^-$ decay is non-trivially correlated with the angular observables, so we also do not include the corresponding observables (\cpp{AI_BKstarmumu} and \cpp{AI_BKstarmumu_zero} in Table \ref{tab:flavourobsBKstar}) in the likelihood function.

Predictions of the branching fractions and forward-backward asymmetries of the inclusive decays $B\to X_s \mu^+\mu^-$ and $B\to X_s \tau^+\tau^-$, corresponding to the last 7 observables of Table \ref{tab:flavourobsWC}, have lower theoretical uncertainties than those of $B^0\to K^{*0}\mu^+\mu^-$. They are however not included in the \flavbit electroweak penguin likelihood, as they provide little additional constraining power when $B\to X_s\gamma$ is already included in a fit --- and only $B\to X_s \ell^+\ell^-$ (where $\ell$ does not distinguish between $e$ and $\mu$) and its forward-backward asymmetry have been measured by BaBar and Belle \cite{Aubert:2004it,Iwasaki:2005sy,Sato:2014pjr,Lees:2013nxa}, with higher uncertainties than measurements of the exclusive modes. We expect to include likelihoods for these observables in a future revision of \flavbit.

\flavbit reads the experimental measurements and correlations, collects them together with the theoretical predictions and uncertainties, and publishes them to the rest of \GB under the capability \cpp{b2sll\_M}.  \flavbit then uses the measurements and correlations to compute the electroweak penguin decay likelihood, which is assigned capability \cpp{b2sll\_LL}.  See Table \ref{tab:likelihoods} for more details.

\subsection{Rare purely leptonic likelihood}

Experimentally, only the decays with muons in the final state have been observed, and therefore give the strongest constraints. For $B_s^0\to \mu^+\mu^-$, we adopt the latest result from LHCb \cite{Aaij:2017vad},
\begin{equation}
{\cal B}(B_s^0\to \mu^+\mu^-)=(3.0\pm 0.6^{+0.3}_{-0.2})\times 10^{-9}\;.
\end{equation}
For $B^0\to \mu^+\mu^-$, we take the results of Ref.~\cite{CMS:2014xfa}, which combines the measurements of the LHCb~\cite{Aaij:2013aka} and CMS experiments~\cite{Chatrchyan:2013bka},
\begin{equation}
{\cal B}(B^0\to \mu^+\mu^-)=(3.9^{+1.6}_{-1.4})\times 10^{-10}\;.
\end{equation}
Experimental correlations between the two decays are negligible \cite{Aaij:2017vad}.

Although the ATLAS collaboration have also recently measured these two branching fractions~\cite{Aaboud:2016ire}, they do not yet report a 3$\sigma$ evidence for these decays. We thus do not include the ATLAS result in \flavbit at this stage. The similar decays $B_{(s)}^0\to e^+e^-$ and $B_{(s)}^0\to \tau^+\tau^-$ have not been measured to date. Only weak upper limits exists in these cases~\cite{Aaltonen:2009vr,Aaij:2017xqt,Aubert:2005qw}, which are currently much less constraining for models of new physics than the muon channels; we therefore do not include them in the \flavbit likelihood.

From the theoretical point of view, $B_{(s)}^0\to \mu^+\mu^-$ decays are rather clean. The theory uncertainty is 10\%, and is dominated by the knowledge of the meson decay constant $f_{B_s}$ \cite{Buras:2012ru}.  This is far smaller than the experimental uncertainty, and therefore has little impact. We also neglect corresponding correlations in the theoretical uncertainties associated with the two decays.

\flavbit reads the experimental measurements and theory errors, collects them together with the theoretical predictions, and publishes them to the rest of \GB as \cpp{b2ll\_M}. It then computes the rare purely leptonic decay likelihood from the measurements and uncertainties, and labels it with capability \cpp{b2ll\_LL}. Table \ref{tab:likelihoods} gives full details.

\subsection{Rare radiative $B$ decay likelihood}

\flavbit includes the average \cite{Misiak:2017bgg} of the measurements of $B\to X_s \gamma$ from BaBar \cite{Aubert:2007my,Lees:2012wg,Lees:2012ym} and Belle \cite{Saito:2014das,Belle:2016ufb} for $E_\gamma > 1.6$\,GeV,
\begin{equation}
{\cal B}(B\to X_s \gamma) = (3.27 \pm 0.14) \times 10^{-4}.
\end{equation}
We adopt a theoretical uncertainty of $7\%$, coming partly from non-perturbative effects \cite{Misiak:2015xwa,Czakon:2015exa}. The corresponding likelihood has capability \cpp{b2sgamma_LL} (Table \ref{tab:likelihoods}), and consists of a direct call to the standard \GB Gaussian likelihood \cite{gambit}.  Note that in general the theoretical calculation from \superiso should be preferred over the corresponding quantity from \FH as input to this likelihood, as the cut employed on the photon energy in \cpp{SI_bsgamma} ($E_\gamma > 1.6$\,GeV -- see Table \ref{tab:flavourobsWC}) is correctly matched to the cut applied in the experimental analysis.

The experimental correlation between $\mathcal{B}(B\to X_s \gamma)$ and the isospin asymmetry of $B\to K^* \gamma$ is not known, though it is expected to be non-negligible given that the event selections overlap.  Because the inclusive branching ratio of $B\to X_s \gamma$ has a smaller theoretical uncertainty, we include $\mathcal{B}(B\to X_s \gamma)$ but not $\Delta_0$ in the likelihood function.

\subsection{$B$ meson mass asymmetry likelihood}

The parameters $\Delta M_s$ and $\Delta M_d$ have been precisely measured~\cite{Amhis:2016xyh}:
\begin{eqnarray}
\Delta M_d &=& 0.5064 \pm 0.0019\; \text{ps}^{-1}\;,\label{exp_deltamd} \\
\Delta M_s &=& 17.757 \pm 0.021\; \text{ps}^{-1}\;. \label{exp_deltams}
\end{eqnarray}
The measurement of $\Delta M_d$ is the average of the results from the DELPHI, ALEPH, L3, OPAL, CDF, D0, BaBar, Belle and LHCb experiments, while the $\Delta M_s$ value is the average of the results from the CDF and LHCb experiments.  The sensitivity of these observables is diluted by the theory uncertainty, which is essentially the same for both SM and BSM predictions, as it is dominated by lattice calculations of non-perturbative effects and the uncertainty on the $B$ decay constant $f_B$. The total theoretical uncertainty on $\Delta M_s$, for example, is currently $15\%$~\cite{Artuso:2015swg}.

At present, \flavbit can predict only $\Delta M_s$ (Table \ref{tab:flavourobswoWC}), so the $B$ meson mass asymmetry likelihood simply compares this prediction to Eq.~\ref{exp_deltams}, using a theoretical error of 15\% and the standard \GB Gaussian likelihood function \cite{gambit}.  This likelihood is available via the capability \cpp{deltaMB_LL} (Table \ref{tab:likelihoods}).

\subsection{Other observables}

The $R_{\mu 23}$ average is dominated by the KLOE~\cite{Ambrosino:2009aa} and NA62~\cite{Lazzeroni:2012cx} experiments. While both $R_\mu$ and $R_{\mu 23}$ are implemented as observables in \flavbit, they are not included in the likelihood. For several BSM models, such as the 2HDM, they add negligible additional constraints, particularly when the decay $B^\pm\to \tau \nu_\tau$ is included in the likelihood via \cpp{SL_likelihood}.

\section{Examples}
\label{sec:examples}

Basic examples of how to use \flavbit in a \gambit BSM global fit can be found in any of the canonical \GB SUSY examples in the \term{yaml_files} directory: \term{CMSSM.yaml}, \term{NUHM1.yaml}, \term{NUHM2.yaml} or \term{MSSM7.yaml} \cite{gambit,CMSSM,MSSM}.  In this section, we go through a number of flavour-specific examples, ranging from flavour-only supersymmetric and effective field theory scans with \GB, to an example of how to use \flavbit in standalone mode.

\subsection{Supersymmetric scan}

It is often instructive to consider the impacts of restricted classes of observables on broader global fits.  In \term{yaml_files/FlavBit_CMSSM.yaml}, we give an example of a Constrained MSSM (CMSSM) fit focussing specifically on observables and likelihoods from \flavbit.  This scan varies three dimensionful Lagrangian parameters defined at the GUT scale (the trilinear coupling $A_0$, the universal scalar mass $m_0$ and the universal fermion mass $m_{\frac12}$), the dimensionless ratio of Higgs VEVs at the weak scale ($\tan\beta$), and two SM nuisance parameters ($\alpha_s$ and $m_t$).  The parameters and ranges are shown in Table \ref{tab:param}.

\begin{table}
\begin{center}
\begin{tabular}{l c c c}
\hline
Parameter & Minimum & Maximum & Prior      \\
\hline
$m_0$       & 50\,GeV   & 7\,TeV  & log    \\
$m_\frac12$ & 50\,GeV   & 5\,TeV  & log    \\
$A_0$       & $-$10\,TeV  & 10\,TeV & hybrid \\
$\tan\beta$ & 3         & 70      & flat   \\
\hline
$\alpha_s^{\MSBar}(m_Z)$ & 0.1167 & 0.1203 & flat   \\
$m_{t,\text{pole}}$     & 171.06 & 175.62 & flat   \\
\hline
\end{tabular}
\end{center}
\caption{\label{tab:param} CMSSM parameters varied in the example fit, along with their associated ranges and prior types.  The ``hybrid'' prior on $A_0$ is logarithmic for $|A_0|>100$\,GeV and flat for $|A_0|<100$\,GeV.}
\end{table}

\begin{figure}
\hspace{-5mm}
\includegraphics[width=1.05\columnwidth]{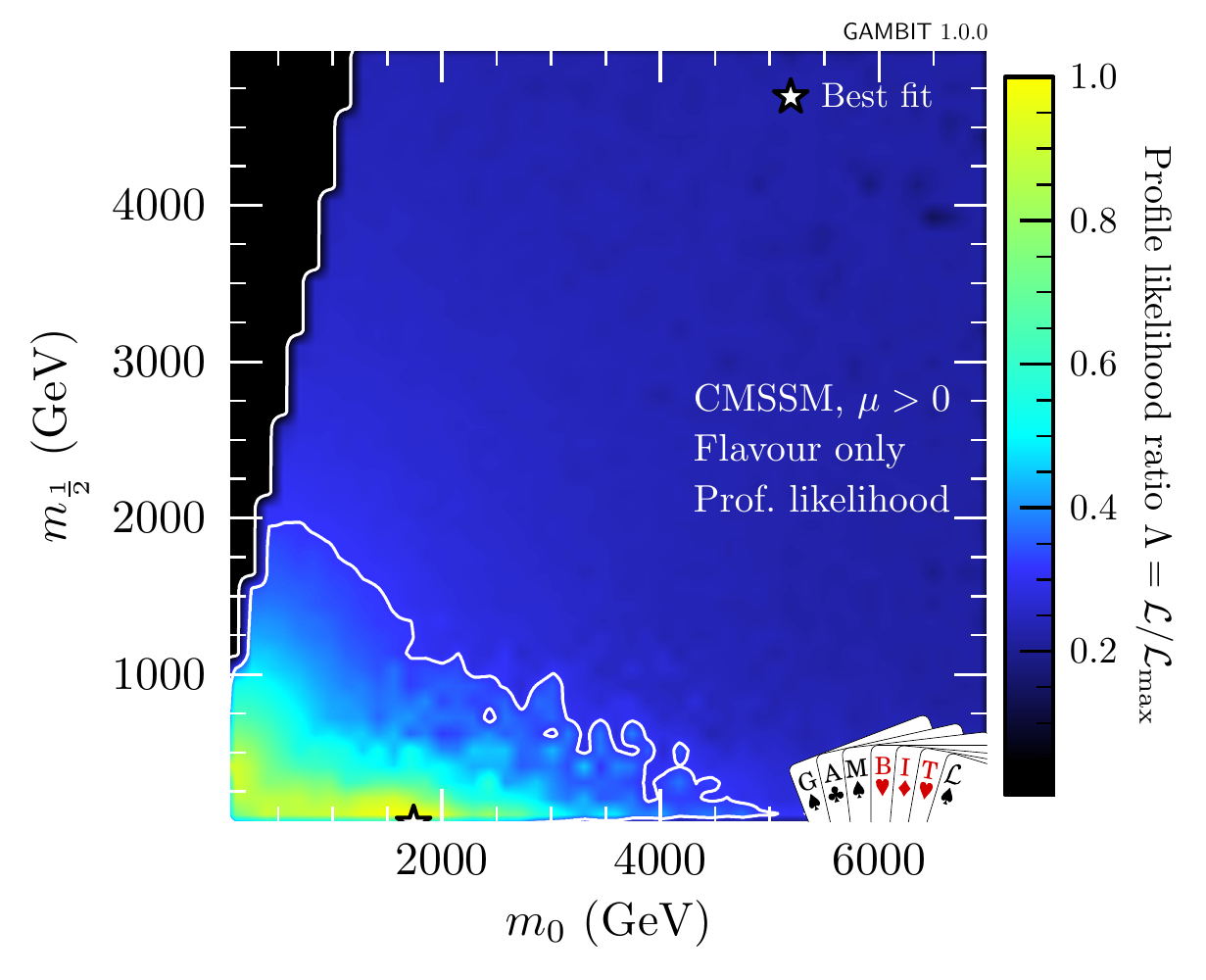}\\
\includegraphics[width=0.9\columnwidth]{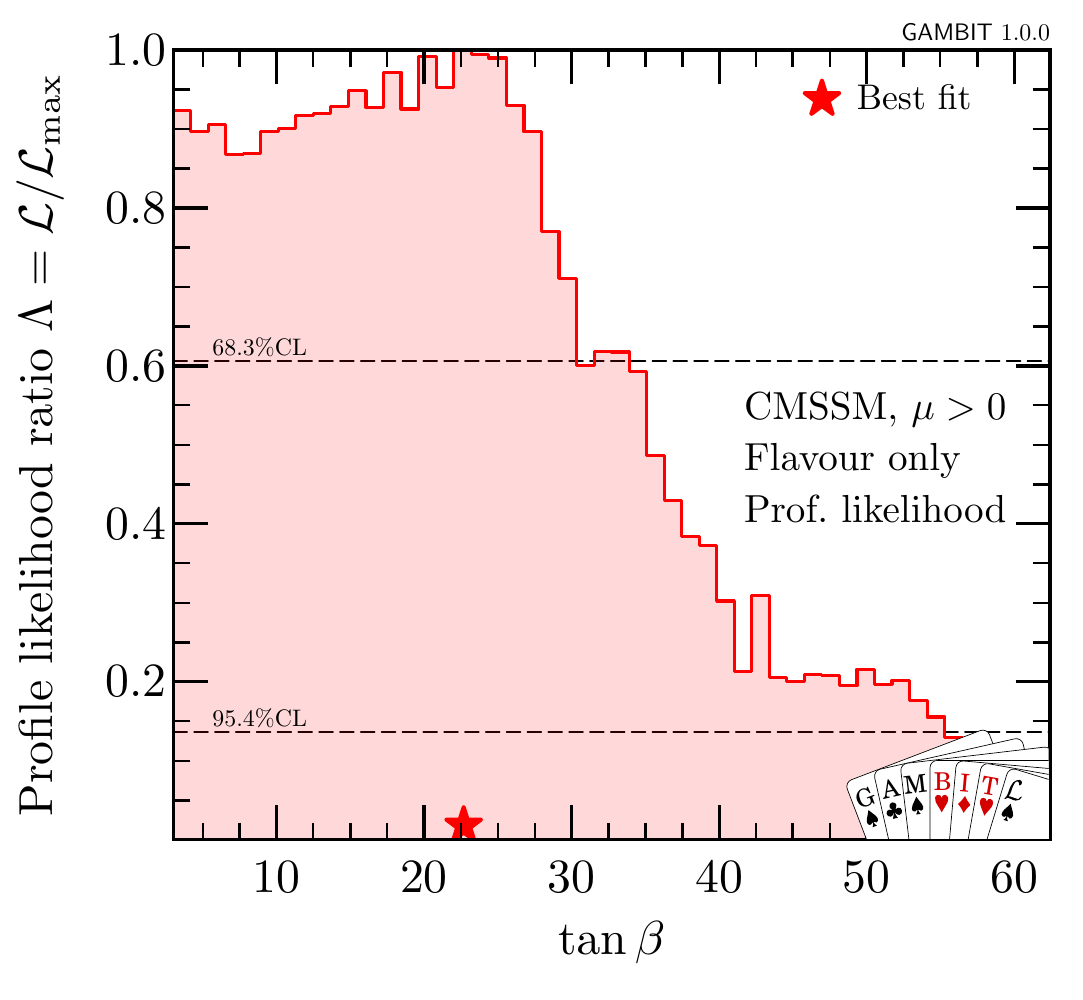}
\caption{2D (upper) and 1D (lower) profile likelihoods of Lagrangian parameters $m_0$, $m_{1/2}$ and $\tan\beta$ in a CMSSM fit including flavour and nuisance likelihoods only.  Stars identify the best fit, and contours indicate 1 and $2\sigma$ confidence regions.  The jagged edge of the $2\sigma$ contour at low $m_0$ and large $m_{1/2}$ is a plotting artefact, caused by interaction of the binning required for plotting and the abruptness of the dropoff of the likelihood in this region (due to the requirement that the lightest supersymmetric particle be a neutralino).}
\label{fig:CMSSM}
\end{figure}

\begin{figure*}
\hspace{-5mm}
\begin{tabular}{c@{}c@{}c}
\includegraphics[height=0.28\textwidth]{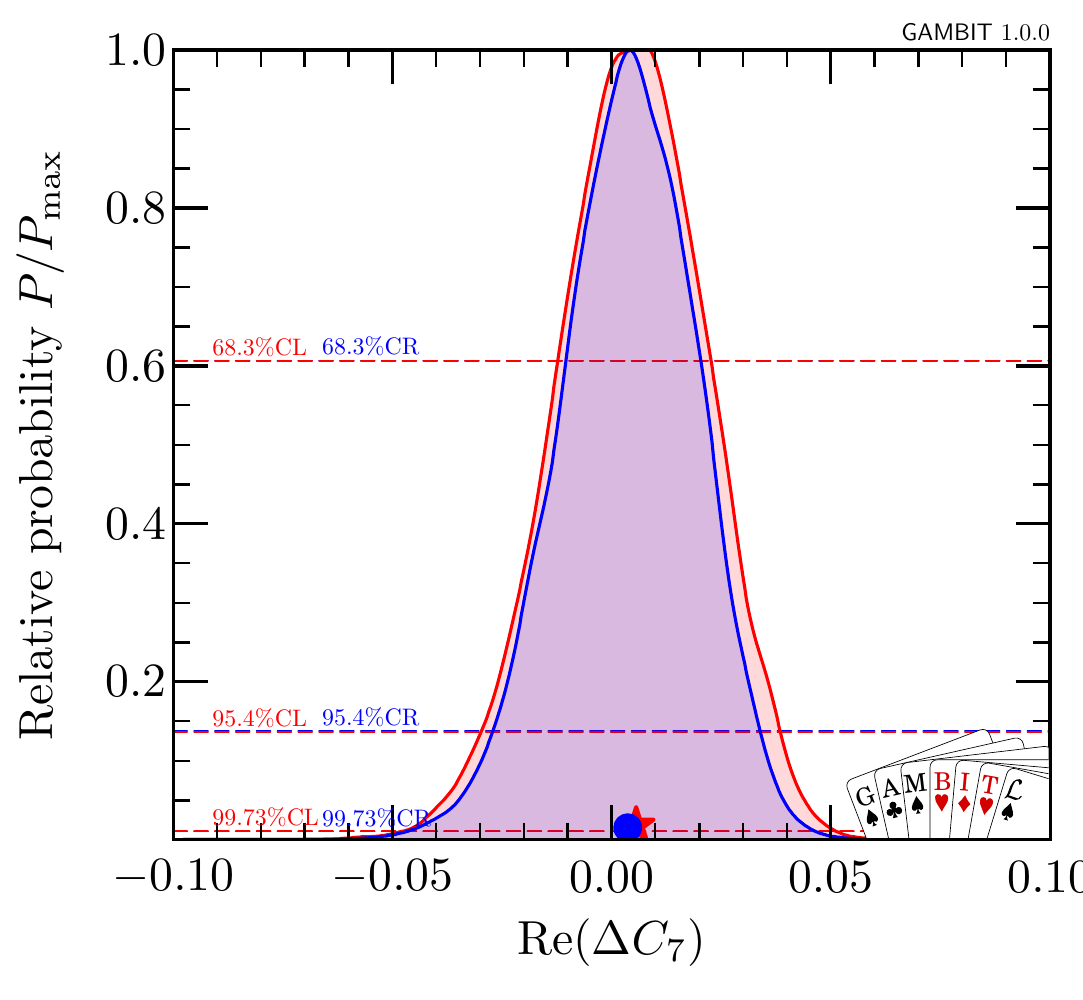}&%
\includegraphics[height=0.28\textwidth]{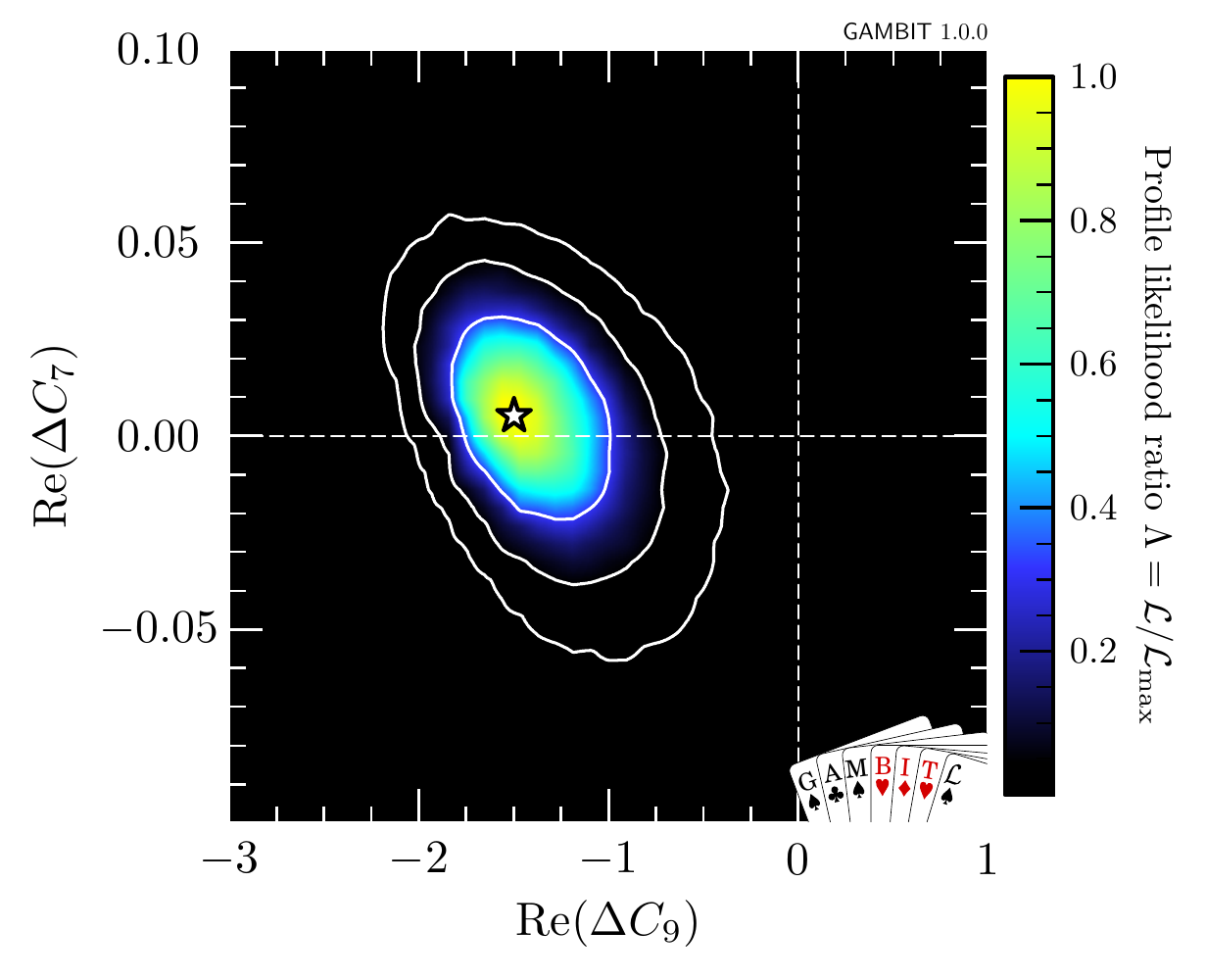}&%
\includegraphics[height=0.28\textwidth]{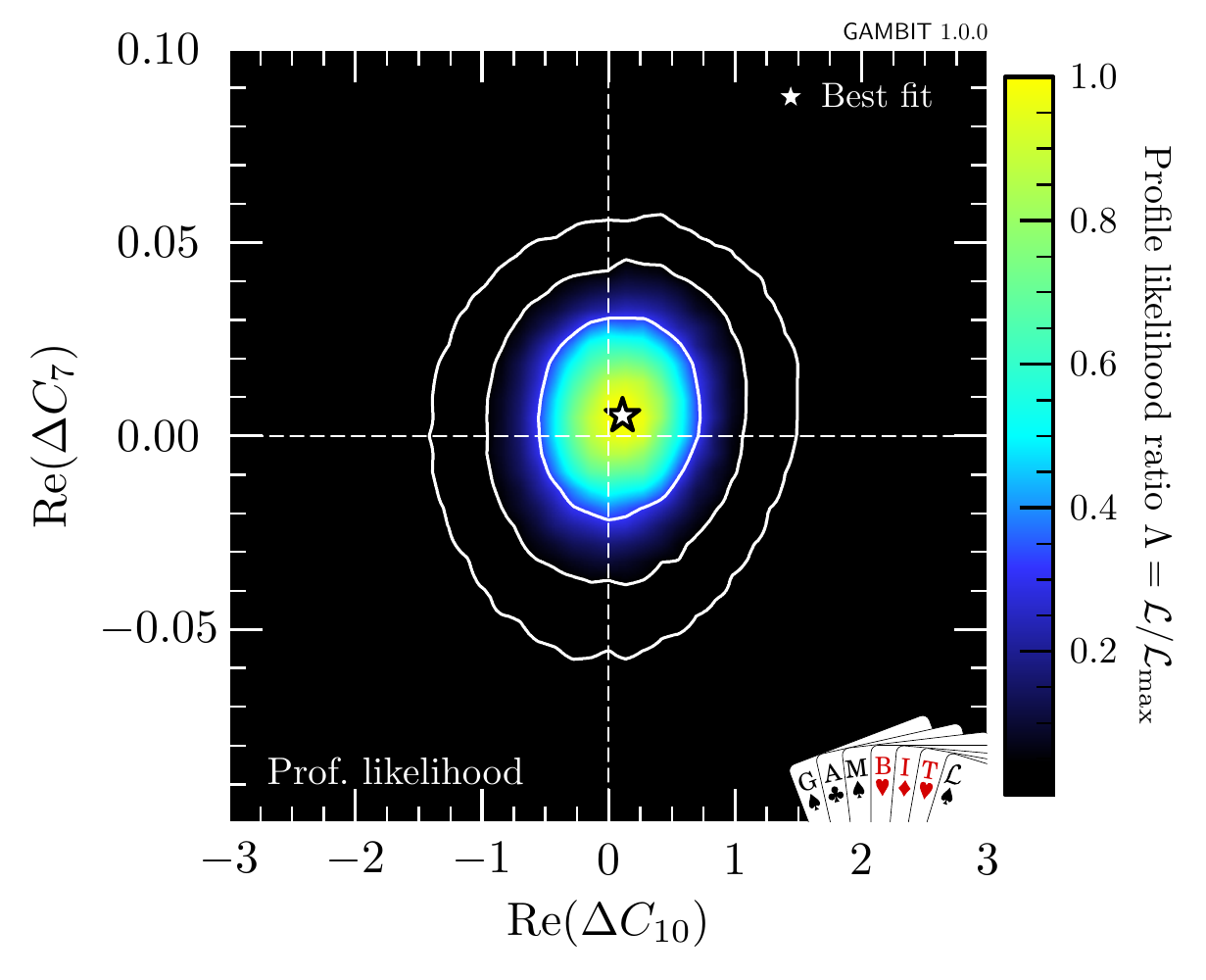}\\
\includegraphics[height=0.28\textwidth]{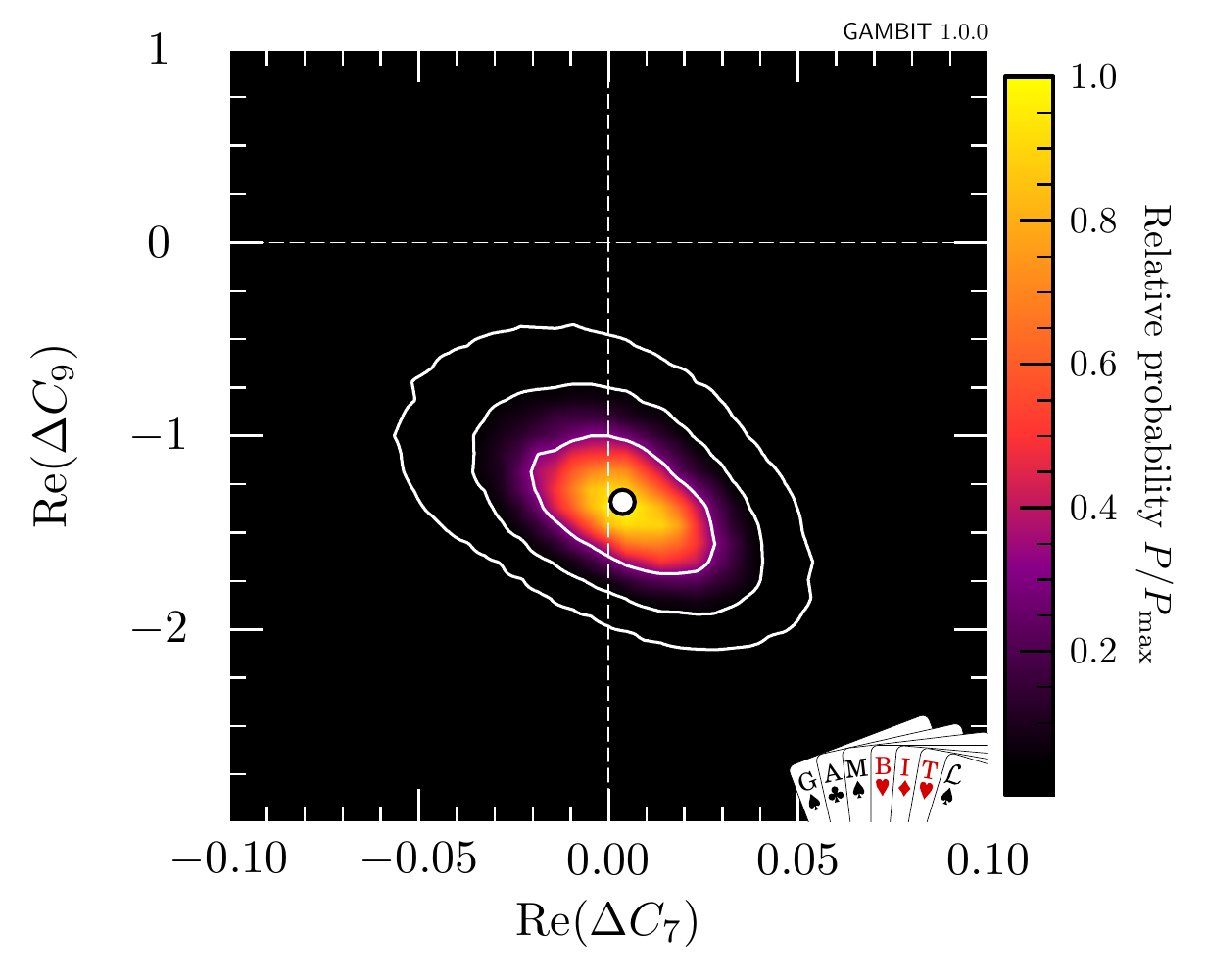}&%
\includegraphics[height=0.28\textwidth]{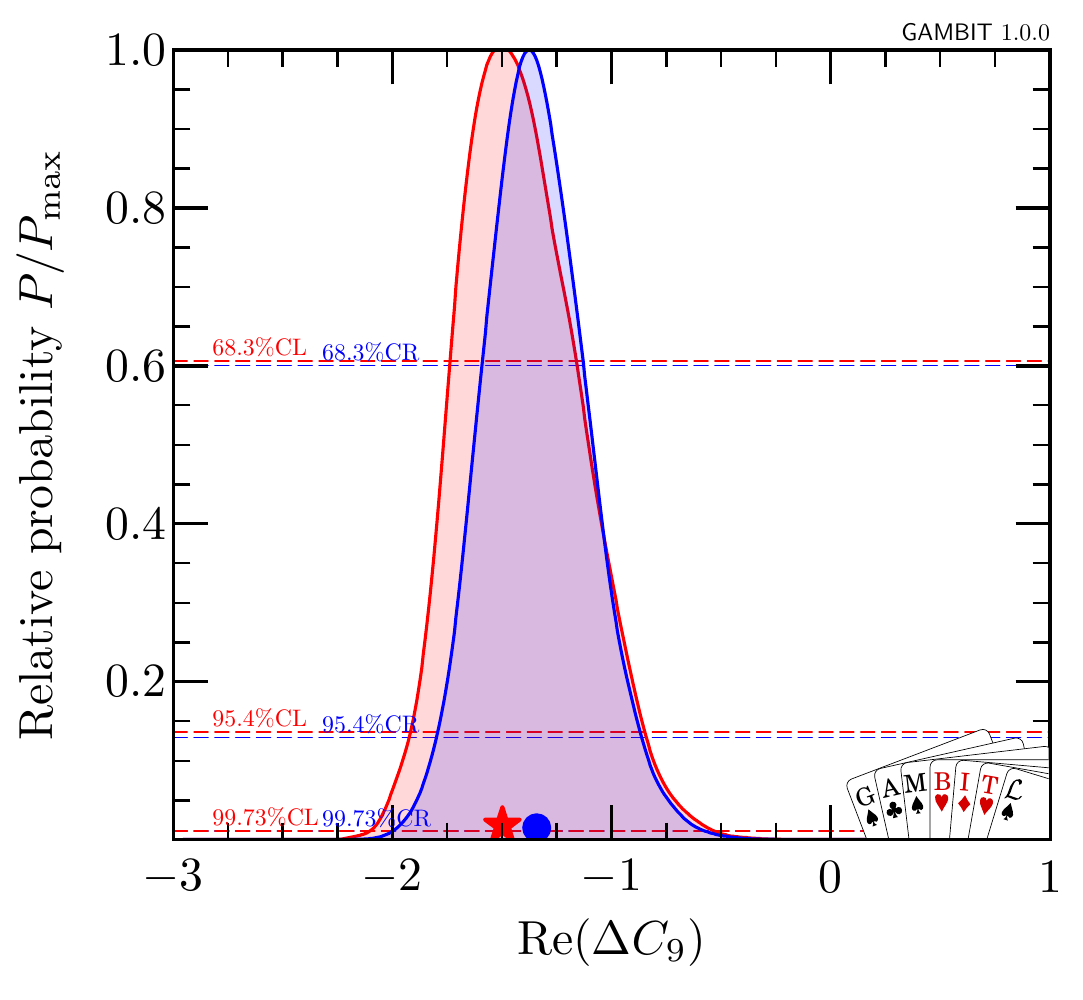}&%
\includegraphics[height=0.28\textwidth]{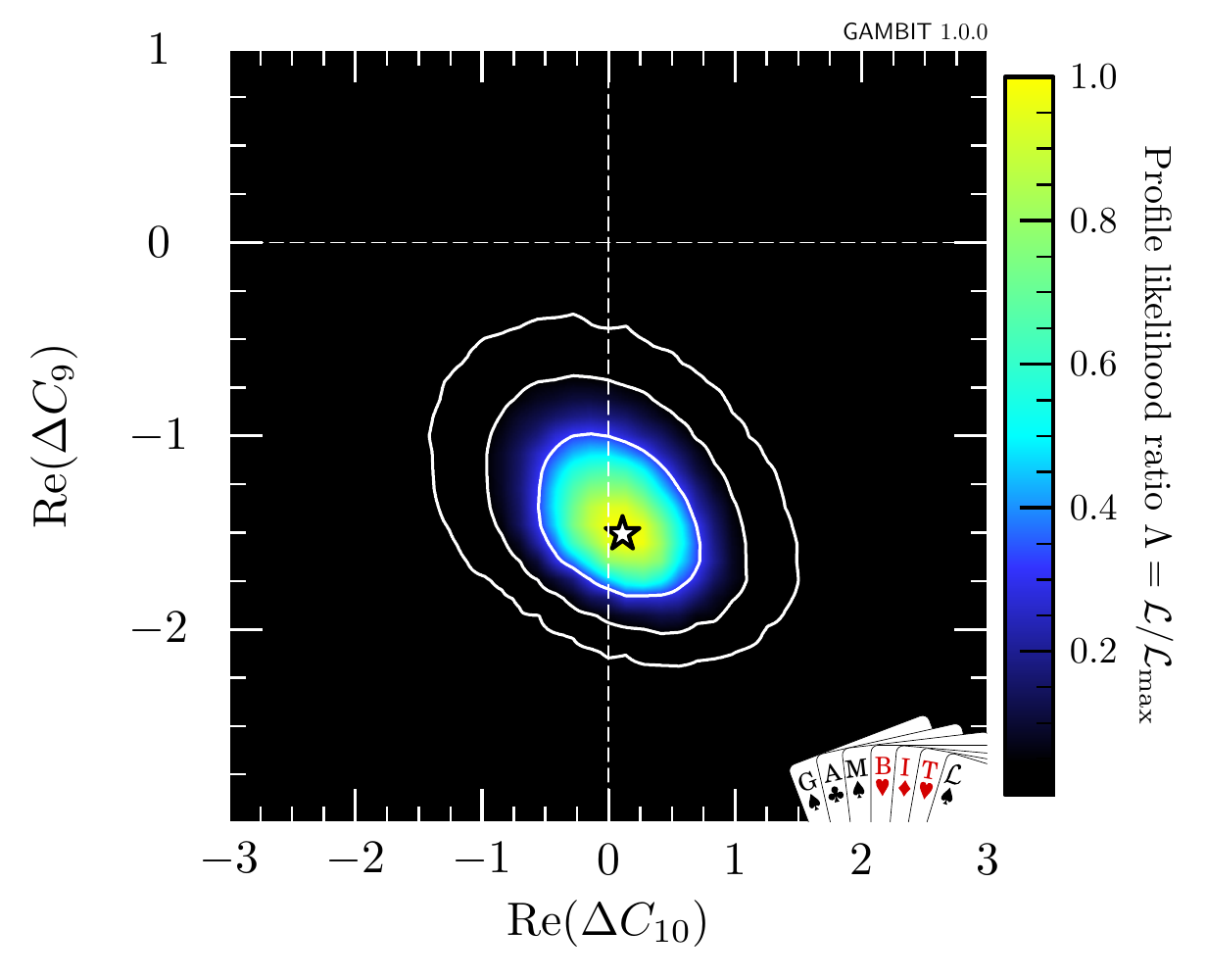}\\
\includegraphics[height=0.28\textwidth]{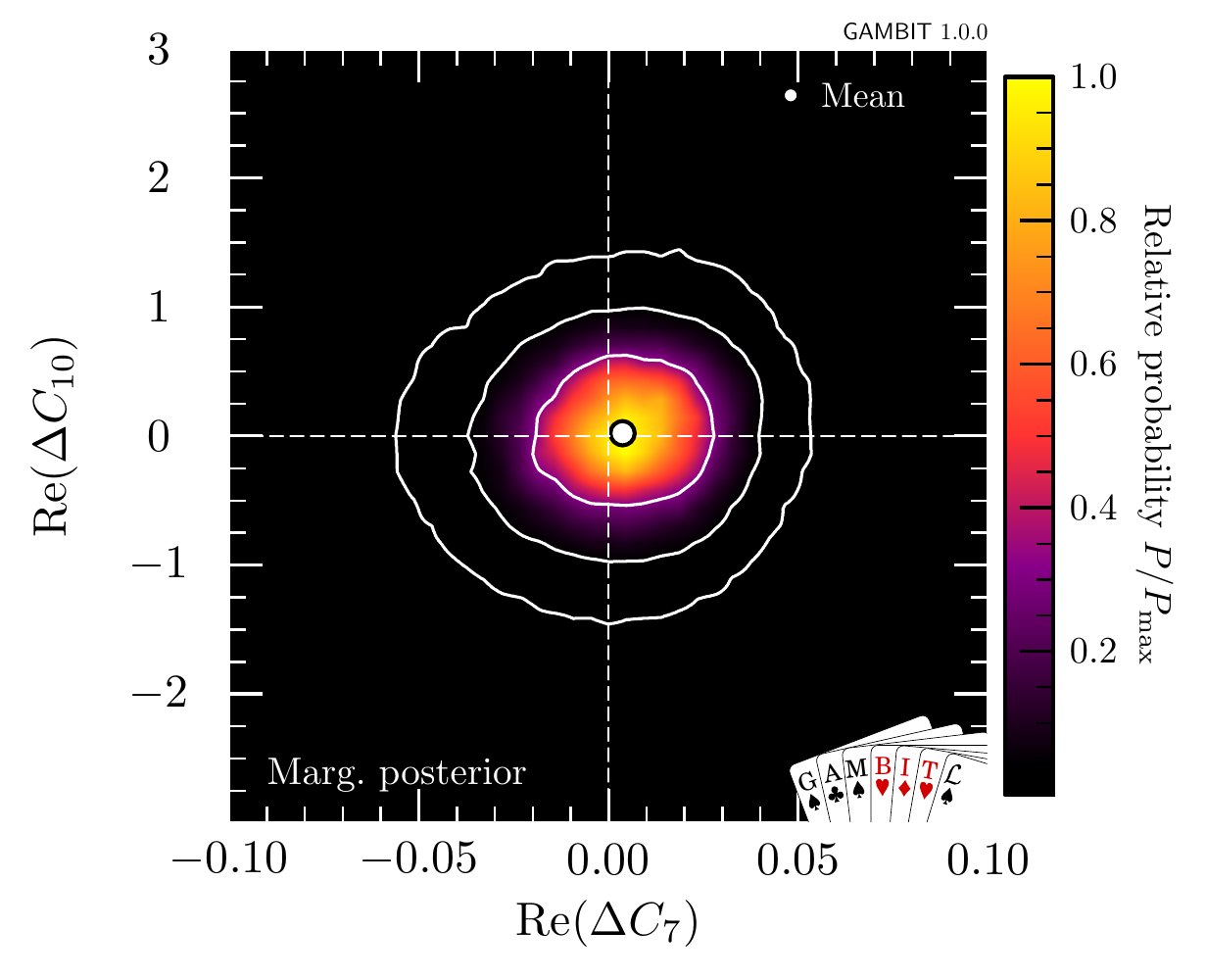}&%
\includegraphics[height=0.28\textwidth]{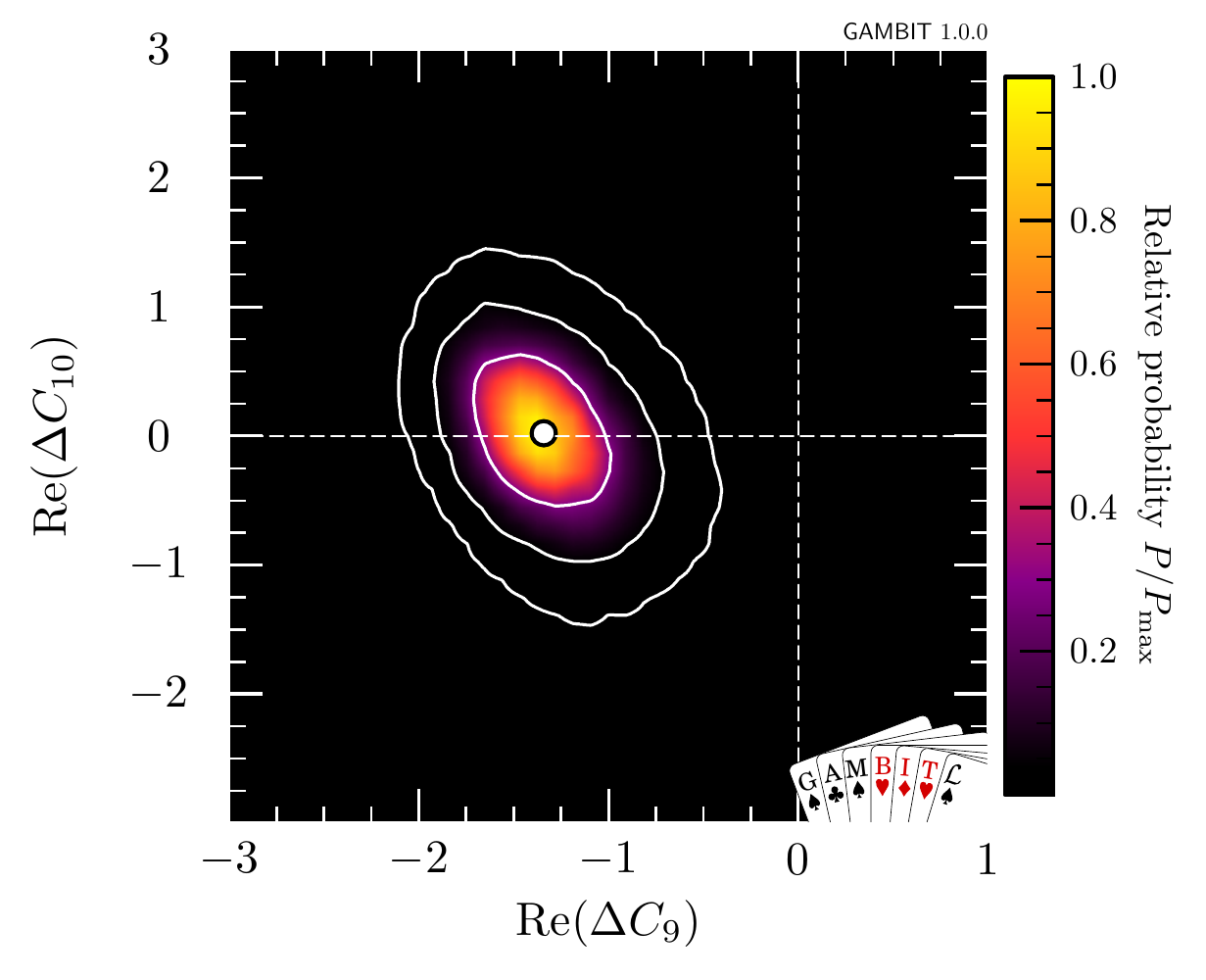}&%
\includegraphics[height=0.28\textwidth]{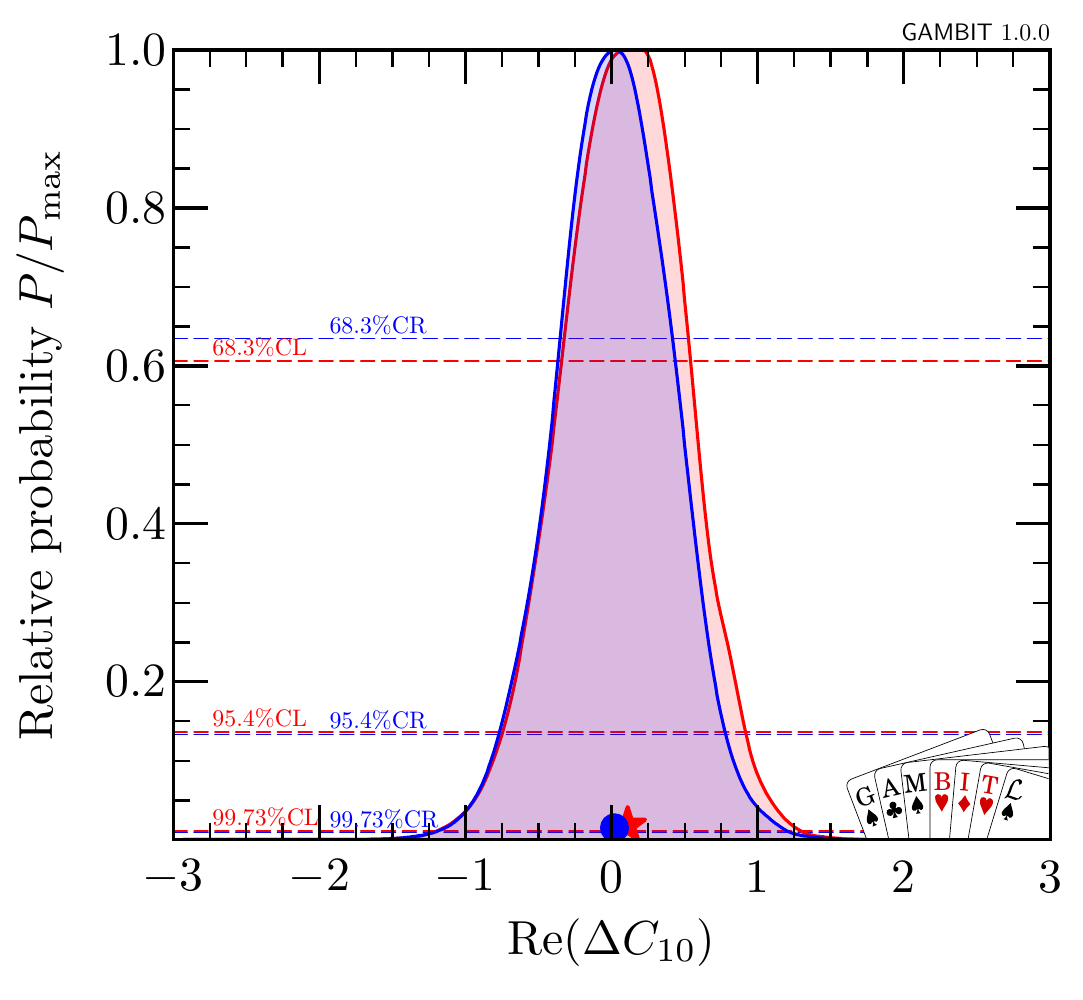}
\end{tabular}
\caption{Profile likelihoods (upper right panels) and posterior probabilities (bottom left panels) from a scan over the real parts of the Wilson coefficients $C_7$, $C_9$ and $C_{10}$, expressed in terms of the offsets $\Delta C_i$ from the SM values.  The central diagonal shows both 1D posterior probabilities (blue) and profile likelihoods (red) for each parameter.  Stars indicate the location of the best fit, filled circles indicate posterior means, and contours correspond to 1, 2, and $3\sigma$ confidence.  The SM prediction lies at the intersection of the dashed lines in the 2D panels. \label{fig:WC}}
\end{figure*}

In this example scan, we include the \flavbit rare leptonic and semileptonic (\cpp{SL_LL}), electroweak penguin (\cpp{b2sll_LL}), rare purely leptonic (\cpp{b2ll_LL}) and rare radiative likelihoods (\cpp{b2sgamma_LL}).  In the interests of speed, numerical stability and comparability to the main CMSSM results presented in Ref.\ \cite{CMSSM}, we do not include the prediction of $\Delta M_s$ from \FH nor the resulting $B$ mass asymmetry likelihood (\cpp{deltaMB_LL}).  We employ nuisance likelihoods from \precisionbit \cite{SDPBit} to constrain $\alpha_s$ and $m_t$.

We focus specifically on the frequentist profile likelihood in this scan, and therefore employ differential evolution to sample the parameter space, as implemented in \diver \cite{ScannerBit}.  Consistent with Ref.\ \cite{CMSSM}, we choose a population of 19200 and a convergence threshold of $10^{-5}$.  Although the profile likelihood is in principle independent of the chosen sampling method and prior, in practice these have an impact on the sampling efficiency and the ability of a scan to uncover more isolated likelihood modes \cite{Akrami09,SBSpike,ScannerBit}.  Our scans employ effectively logarithmic priors on the dimensionful BSM parameters, and flat priors on all other parameters.  The SM parameters are sufficiently well constrained that the prior is irrelevant.  We discuss the impact of the sampling prior on the BSM parameters below.

The resulting scan took approximately 15 minutes to run on 1200 CPU cores, and produced 1.1 million likelihood samples.

The results are shown in Fig.\ \ref{fig:CMSSM}, in terms of the 2D profile likelihood of the sparticle masses $m_0$ and $m_\frac12$, and the 1D profile likelihood of $\tan\beta$.  The flavour likelihoods have the most impact at large $\tan\beta$, as has been extensively pointed out in the literature (e.g.\ \cite{Arbey:2012ax, Mahmoudi:2014mja}).  The 2D figure shows a weak preference (at the 1--2$\sigma$ level) for lower sparticle masses.  At first glance this may seem surprising, given the lack of hints for SUSY, the fact that the likelihood at large $m_0$ and $m_\frac12$ essentially recovers the SM result, and the resulting tendency of $b\to s\gamma$ to drive SUSY fits to larger masses to avoid spoiling the good agreement between the SM prediction and the observed value of $\mathcal{B}(B\to X_s\gamma)$.  Indeed, the likelihood improvement at low mass is driven entirely by the angular analysis of $B^0\to K^*\mu^+\mu^-$ decays, with the fit attempting to account for the deviation from the SM prediction in this channel by making the new states light and boosting the (generally small) SUSY contributions as much as possible.  This effect is rather small, providing an improvement in the likelihood contribution from $B^0\to K^*\mu^+\mu^-$ (\cpp{b2sll_likelihood}) of $\Delta\ln\mathcal{L}=3.4$ relative to the SM.  This improvement is mostly counteracted by a corresponding decrease of $\Delta\ln\mathcal{L}=-2.0$ in the likelihood associated with $\mathcal{B}(B\to X_s\gamma)$ (\cpp{b2sgamma_likelihood}).

\subsection{Wilson coefficient fit}

As a more advanced example, we carry out a joint fit to the real parts of the $C_7$, $C_9$ and $C_{10}$ effective couplings of Eq.\ \ref{physical_basis}, expressed in terms of offsets from their SM values $\Delta C_i \equiv C_i - C_{i,\text{SM}}$. The \YAML file for this scan can be found at \term{yaml_files/WC.yaml}.

In this example, we use the electroweak penguin likelihood (\cpp{b2sll_likelihood}), the rare purely leptonic decay likelihood (\cpp{b2ll_likelihood}) and the rare radiative decay likelihood (\cpp{b2sgamma_likelihood}). The other two  likelihood functions available in \flavbit (based on the $B$ meson mass asymmetry and tree-level leptonic and semi-leptonic decays) have no dependence on the three Wilson coefficients that we vary.  We also scan over the \MSbar $b$ quark mass and the strong coupling as nuisance parameters, computing associated nuisance likelihoods with \precisionbit \cite{SDPBit}.  We sample the parameter space with nested sampling \cite{Skilling04,MultiNest}, using 20\,000 live points and a tolerance of 0.1; see Ref.\ \cite{ScannerBit} for details of the scanning setup and sampling algorithm.

The results of this scan are shown in Fig~\ref{fig:WC}.  Here we show both Bayesian posterior probabilities (lower left panels) and frequentist profile likelihoods (upper right panels), which are in rather close agreement.  The small offset between the peaks of the posterior and the profile likelihood in $\Delta C_9$ is a volume effect, reflecting the fact that the posterior is slightly broader in $C_7$ and $C_{10}$ at values below the best-fit $\Delta C_9$ than above it.  The results show a $>3\sigma$ preference for a negative offset to the muonic version of the $C_9$ Wilson coefficient compared to the SM, consistent with recent results from other groups \cite{Hurth:2014vma,Altmannshofer:2017fio,Descotes-Genon:2015uva}. These are largely driven by the $B^0\to K^*\mu^+\mu^-$ angular observables, with the corresponding component of the best-fit likelihood improved by $\Delta\ln\mathcal{L}=13.2$ with respect to the SM, and $\Delta\ln\mathcal{L}=9.8$ compared to the CMSSM.  We can also see that $C_7$ is strongly constrained by $b \to s \gamma$ decays, to within $+0.04$/$-0.03$ of its SM value.

\subsection{\flavbit standalone example}

\gambit modules can also be called directly from other codes as libraries, without actually needing to use \gambit itself.  To do this, the calling code must specify the physics model and parameter set to be used, the module and backend functions to be run, and any required options. The calling code is responsible for resolving the dependencies and backend requirements of each module function; this is typically done ``by hand'' by the author of the calling code, using simple \gambit utility functions to hardcode the links between the chosen module and backend functions.  More details of using \GB modules in this so-called `standalone mode' can be found in Ref.\ \cite{gambit}.

An annotated driver program for calling \flavbit from outside the \gambit framework can be found in \term{FlavBit/examples/FlavBit\_standalone\_example.cpp}.  As input, this program takes an SLHA file corresponding to the output of a spectrum generator (i.e.\ containing pole masses, \DRbar parameters, etc).  The name of this file can be given as a command-line argument.  The program then calculates the full menu of \flavbit observables using \superiso \textsf{3.6} and \FH \textsf{2.11.3}, and uses them to calculate the five independent \flavbit likelihoods.  Much of this short program is dedicated to resolving module function dependencies and backend requirements.  This includes defining a local function that creates a \gambit\ \cpp{Spectrum} object from the input SLHA file, and others that fulfil the dependencies of \cpp{SI_fill} on the widths of the $Z$ and $W$ bosons.

If the user does not give the name of an input SLHA file when invoking the standalone example, it will read a default file given in the line
\begin{lstcpp}
std::string infile("FlavBit/data/example.slha");
\end{lstcpp}
The likelihoods are retrieved in the lines
\begin{lstcpp}
loglike = b2ll_likelihood(0);
loglike = b2sll_likelihood(0);
loglike = SL_likelihood(0);
loglike = b2sgamma_likelihood(0);
loglike = deltaMs_likelihood(0);
\end{lstcpp}
and can be combined or used for further analysis as the user requires.

The values of the observables, as used by the likelihoods, can be obtained directly from the respective observable functions in a similar manner, e.g.
\begin{lstcpp}
double bsg = SI_bsgamma(0);
double Btaunu = SI_Btaunu(0);
\end{lstcpp}
and so on for all observables in Tables~\ref{tab:flavourobswoWC}, \ref{tab:flavourobsWC} and \ref{tab:flavourobsBKstar}.

\section{Conclusions}
\label{sec:summary}

In this paper we have described \flavbit, the flavour physics module of the public global-fitting framework \GB.  \flavbit provides calculations of a wide range of observables in flavour physics, ranging from tree-level decays of $B$ and $D$ mesons, to electroweak penguin decays, rare purely leptonic $B$ decays, $b\to s\gamma$ transitions, neutral meson oscillations, kaon and pion decays, and various isospin and forward-backward asymmetries.  These are so far implemented for supersymmetric and effective field theories, with the list of available theories expected to grow rapidly.  \flavbit also features detailed experimental data, uncertainties, correlations and likelihood functions for tree-level leptonic and semileptonic, electroweak penguin, rare purely leptonic and $B\to X_s\gamma$ decays, as well as for the $B^0_s$--$\bar{B}^0_s$ mass difference.

We gave a number of interesting examples of \flavbit in action.  These include a standalone example program that runs \flavbit without \GB, in order to compute flavour observables in supersymmetry from an input SLHA file.  We carried out an example supersymmetric flavour fit with \flavbit in \GB, illustrating the impacts of its likelihoods.  Finally, we performed a fit to a number of observables in the context of an effective theory of flavour, demonstrating about a $4\sigma$ preference from combined experimental data for an approximately $25\%$ deficit in the (muonic) $C_9$ Wilson coefficient, compared to the Standard Model prediction.

The \flavbit source code can be freely downloaded from \href{http://gambit.hepforge.org}{gambit.hepforge.org}, either as part of \GB, or as a standalone package.

\begin{acknowledgements}
We thank our colleagues within \GB for many helpful discussions.
\gambitacknos
\end{acknowledgements}

\appendix

\startglossary
\gitem{backend}\input{"glossary/backend.glossentry"}
\gitem{backend function}\input{"glossary/backend_function.glossentry"}
\gitem{backend requirement}\input{"glossary/backend_requirement.glossentry"}
\gitem{backend variable}\input{"glossary/backend_variable.glossentry"}
\newcommand{\seecompdatabase}{see Sec.\ 10.7 of Ref.\ \cite{gambit}}
\gitem{capability}\input{"glossary/capability.glossentry"}
\gitem{dependency}\input{"glossary/dependency.glossentry"}
\gitem{dependency resolution}\input{"glossary/dependency_resolution.glossentry"}
\gitem{frontend}\input{"glossary/frontend.glossentry"}
\gitem{frontend header}\input{"glossary/frontend_header.glossentry"}
\gitem{module}\input{"glossary/module.glossentry"}
\gitem{module function}\input{"glossary/module_function.glossentry"}
\gitem{physics module}\input{"glossary/physics_module.glossentry"}
\gitem{rollcall header}\input{"glossary/rollcall_header.glossentry"}
\gitem{type}\input{"glossary/type.glossentry"}
\finishglossary

\bibliography{R1}

\end{document}